# Microstructure evolution under the space-time variational solidification conditions in a melt pool: A multi-scale simulation study

Fengyi Yu[a], Yanhong Wei[b], Qiaodan Hu[a,*], Jianguo Li[a]

[a] Shanghai Key Laboratory of Materials Laser Processing and Modification, School of Materials Science and Engineering, Shanghai Jiao Tong University, Shanghai 200240, PR China

[b] College of Material Science and Technology, Nanjing University of Aeronautics and Astronautics, Nanjing 210016, PR China

[*] Corresponding author, E-mail address: qdhu@sjtu.edu.cn, Tel: 0086-21-54744246

**Abstract**

The properties of welded components are dominated by the microstructure evolution in the pool, where the solidification conditions are space-time variational. To represent the variational solidification conditions in the pool, the multi-scale simulation is carried out in this paper, combining the microscopic Phase-Field (PF) equations with the macroscopic thermal processes. Firstly, two different models, the GR model and TF model, are employed to simulate the single crystal solidification at a local region of the pool. Results suggest the TF model is more suitable to reflect the variational conditions than the GR model. Secondly, the single-crystal solidification and poly-crystal solidification at the whole region of the pool are performed through the TF model. The results demonstrate the space-time variabilities of the solidification conditions across the melt pool. Meanwhile, the variational conditions affect the microstructure evolution significantly, including the onset of initial instability at the epitaxial growth stage and the directional evolutions of the converging grain boundaries (GBs) and diverging GBs at the competitive growth stage. Moreover, the formation of axial grain structures is observed, which can be regarded as the competition between the grains along the axial direction and radial direction. This study indicates the necessity of considering the variational conditions in a pool. Meanwhile, the PF model can simulate microstructure evolution under the variational conditions accurately, which has a great potential for investigating solidification dynamics in the melt pool.

**Keywords**

Microstructure evolution; Space-time variational conditions; Melt pool solidification; Multi-scale simulation; Phase-field model.

## 1. Introduction

The properties of fusion-welded components are determined by the microstructures of the Fusion Zone (FZ), dominated by the melt pool solidification [1-3]. In past decades, the microstructure evolution in melt pool solidification has been investigated by various approaches, including experimental observations, analytical models and numerical simulations [3-5]. The numerical simulations can reproduce the dynamical evolution of solidification, by calculating out the space-time distributions of various physical fields, which have been applied for the investigations of microstructure evolution in the melt pool. In recent years, there has been an increasing interest in the numerical simulation of melt pool solidification [6-16]. The ultimate goal is to predict the microstructures after melt pool solidifying. It should be noted the heat source moving causes the temperature field to vary considerably across the pool, making the temperature gradient G and pulling speed R space-time variational. The space-time variabilities of G and R include the continuously changing G and R with time, as well as the varying G and R from fusion line to centerline at the same time. Compared with directional solidification (with the constant G and R), melt pool solidification (with the variational G and R) has specific characteristics in the microstructure evolution [2,3]. To predict microstructures in the melt pool accurately, the space-time variational solidification conditions should be represented.

To represent the space-time variational conditions in the pool, the simulation method needs to cross different scales, i.e., the multi-scale simulation. Due to different characteristics of physical processes at the different scales, the multi-scale simulation has been a long standing computational challenge. Researchers used various methods to accomplish the multi-scale simulation of melt pool solidification. Nguyen et al. [8] developed a Finite Element (FE) multi-scale model, considering the nucleation and solute diffusion, as well as the transports of the liquid and solid. The simulations indicate the impact of solid transport on formation of macro-segregation, demonstrating the ability of the FE method simulating solute segregation caused by thermo-mechanical deformation in material processing. As for microstructure evolution, Wei et al. [9] used a Monte Carlo (MC) model to predict morphological evolution of grains, considering the heat transfer and liquid flow. Then the orientation evolution of columnar grains and Columnar to Equiaxed Transition (CET) were simulated out and validated by the experimental observations, indicating the ability of the MC method predicting the grain evolution in the pool. Rodgers et al. [10] used a MC model to simulate melting and solidification in the FZ, as well as solid-state phase transformation in the Heat Affected Zone (HAZ), in the pulsed power weld. The Bézier curves were used to represent a wide range of melt pool shapes, from narrow and deep to wide and shallow, corresponding to the variational solidification conditions. The results show the microstructures vary considerably with the shape of pool, as well as the location of pool. Moreover, even small deviations in weld parameters lead to large variations in microstructures. The study indicates the importance of considering the variational solidification conditions. To investigate the effect of solidification conditions on microstructures, Han et al. [11] developed a macro-micro model to predict the solidification structures in a melt pool of Gas Tungsten Arc Weld (GTAW). The macro model predicted the processes of heat and mass transfer, based on which a Cellular Automaton (CA) model was developed to simulate the solidification evolution in the pool. The results show the solidification parameters G and R vary with the location of pool, from fusion line to centerline. The solidification evolutions at different locations also follow different formation mechanisms, including the columnar grain formation along fusion line and the equiaxed grain formation near centerline. The study indicates the variational G and R affect microstructure evolution in the pool significantly. Furthermore, to study the effect of microstructures on the properties of welded parts, Zareie Rajani et al. [13] developed a multi-scale multi-physics model to predict hot crack formation in the GTAW of aluminum alloy, considering various factors on the Hot Crack Susceptibility (HCS), including the response of base metal, solidification contraction, external deformation and fluid flow. Then an analytical model was developed to represent the effect of each factor on the HCS intuitively. The study reveals the importance of the microstrcutres on predicting the properties of welded components. In conclusion, from the application viewpoint of the multi-scale simulation, different methods are applied for different purposes. The FE method could simulate the transfer processes of heat and mass and macro-segregation. The MC method can predict the grain evolution in relatively large domain, in the FZ and FAZ. The CA method can reproduce the evolution of structures, considering the variational conditions and the stochastic factors of nucleation and growth, at the mesoscopic scale. The analytical model can provide an intuitionistic expression of the relative parameters. However, for the precise analysis of solidification dynamics and the microstructure prediction, the previously mentioned methods could hardly meet the accuracy requirement.

The Phase-Field (PF) method, avoiding the shape error caused by tracking interface, could accurately simulate the microstructure evolution and capture the detailed information of interface [17-20]. For alloy solidification, a quantitative PF model was developed by Karma and co-workers [21-23]. The PF model has been widely applied for investigating solidification dynamics at the mesoscopic scale and considerable achievements have been made [24-27]. Moreover, since it formulates unified governing equations across the whole domain, the PF model can be extended for considering other physical fields [28]. This characteristic reveals the potential of the PF model coupling macroscopic temperature field, which means the variational solidification conditions can be represented. As for weld, there has been a growing number of publications focusing on solidification dynamics via the quantitative PF model, from steady-state conditions [29,30] to transient conditions [31,32], from single-crystal solidification [30,31,33] to poly-crystal solidification [15,16,34]. The consistency between the simulations and experimental observations indicates the capability of the quantitative PF model predicting the microstructures in the pool accurately. Based on the PF results, researchers investigated the solidification dynamics under variational conditions in the pool, including the effects of surface energy and grain size on the morphological evolution [15], as well as the effects of tip-splitting and sidebranches on dendrite arm spacing [16], etc. However, the existing researches have suffered from a lack of systematic study on the effects of variational conditions on the microstructures. As mentioned before, the space-time variational G and R in the pool affect the microstructure evolution significantly [2,3]. To investigate the evolution mechanism of melt pool solidification accurately, the influences of variational conditions on the microstructures need to be clarified.

In this paper, the microstructure evolutions under the variational conditions in melt pool solidification are carried out by the multi-scale simulation, connecting the microscopic PF equations with the macroscopic thermal parameters. Firstly, we use two different models, the GR model and TF model, to simulate the single-crystal solidification at a local region of the pool. The results indicate the necessity of considering the space-time variational conditions, especially at the late stage of solidification. Moreover, the TF model can reflect the variational conditions better than the GR model. Secondly, the single-crystal solidification at the whole region is performed via the TF model. The evolutions of the parameters G, R, G*R and G/R are obtained, indicating the space-time variabilities of solidification conditions across the pool. Based on the single-crystal solidification at the whole region, the effects of crystallographic parameters and solidification parameters on the microstructures are discussed. Moreover, reproduce melt pool solidification more realistically, the poly-crystal solidification at the whole region is carried out via the TF model. Then the influences of variational conditions and Grain Boundary (GB) on solidification evolution are discussed, including the onset of initial instability in epitaxial growth, the evolutions of converging GBs and diverging GBs in competitive growth, and the formation of axial grain structures along the centerline. Once again, these special characteristics indicate the necessity of considering the space-time variational conditions in the melt pool solidification.

## 2. Models and methodology

### 2.1. Quantitative PF model of alloy solidification

We used the quantitative PF model for alloy solidification [23,35], with solute diffusion in the solid [36,37]. The detailed derivations and validations of the PF model could be found in literatures [23,35-37]. Here we just present the equations describing the evolutions of phase field and solute field.

For the phase field, a set of scalar variables $\phi_i(\mathbf{r}, t)$ is introduced to describe the phase states at the given location and time. Specifically, the subscript $i$ reflects the order of the phase. $\phi_i(\mathbf{r}, t) = 1$ means, at the given location and time, only the i-th phase is in solid state, while all the other phases are in liquid state, i.e., $\phi_j(\mathbf{r}, t) = -1$, where $j \neq i$. The values of $\phi_i(\mathbf{r}, t)$ smoothly cross the solid/liquid (S/L) interface and the GBs. It should be noted that the number of phases in solid state cannot be greater than one at the given location.

For the solute field, the composition $c(\mathbf{r}, t)$ is represented via the supersaturation field $U(\mathbf{r}, t)$:

$$U = \frac{1}{1-k}\left(\frac{2kc/c_\infty}{1+k-(1-k)\cdot\psi} - 1\right) \quad (1)$$

where k means the solute partition coefficient, $c_\infty$ is the average solute concentration. $\psi = -1+\sum(\phi_i+1)$ is an interpolation function describing the phase state. The expression of $\psi$ assures the value changes from -1 to 1.

For alloy solidification, the introduction of "Anti-Trapping Current" (ATC) could recover the local equilibrium at the S/L interface. Moreover, the ATC term could eliminate the spurious effects when the interface width is larger than the capillary length [22,23]. The ATC term for poly-crystal with solute diffusion in the solid is expressed by [36,37]:

$$\vec{j}_{at} = -\frac{1-k\cdot D_S/D_L}{2\sqrt{2}}[1+(1-k)U]\frac{\partial\psi}{\partial t}\frac{\vec{\nabla}\psi}{|\nabla\psi|} \quad (2)$$

where $D_S$ and $D_L$ represent the diffusion coefficients in solid and liquid phases, respectively. $\partial\psi/\partial t$ means the rate of solidification, $\nabla\psi/|\nabla\psi|$ is the unit length along the normal direction of interface.

For Al-Cu alloy with cubic crystal anisotropy, a four-fold anisotropy function is adopted:

$$a_s(\hat{n}) \equiv a_s(\theta_i + \theta_i^0) = 1 + \varepsilon_4 \cos 4(\theta_i + \theta_i^0) \quad (3)$$

where $\varepsilon_4$ is the anisotropy strength, $\theta_i$ the angle between the normal direction of S/L interface and the y-axis, $\theta_i^0$ is the intersection angle between the Preferred Crystallographic Orientation (PCO) of the i-th phase and the y-axis.

Finally, the governing equations of phase field and supersaturation field could be expressed by:

$$a_s^2(\hat{n})\left[1-\frac{k(T-T_0)}{|m|c_\infty}\right]\frac{\partial \phi_i}{\partial t} =$$
$$\nabla \cdot \left[a_s^2(\hat{n})\vec{\nabla}\phi_i\right] - \partial_x\left(a_s(\hat{n})\cdot a_s^{'}(\hat{n})\cdot \partial_y\phi_i\right) + \partial_y\left(a_s(\hat{n})\cdot a_s^{'}(\hat{n})\cdot \partial_x\phi_i\right) \quad (4)$$
$$+\phi_i(1-\phi_i^2) - \lambda(1-\phi_i^2)^2\left[U+\frac{k(T-T_0)}{|m|c_\infty(1-k)}\right] - \omega\frac{(1+\phi_i)}{2}\sum_{j\neq i}\left(\frac{1+\phi_j}{2}\right)^2$$

$$\left(\frac{1+k}{2} - \frac{1-k}{2}\psi\right)\frac{\partial U}{\partial t} =$$
$$\nabla \cdot \left[\overline{D}_L \cdot q(\psi) \cdot \vec{\nabla}U - \vec{j}_{at}\right] + \frac{1}{2}\left[1+(1-k)U\right]\frac{\partial \psi}{\partial t} \quad (5)$$

where,

$$\overline{D}_L = D_L / (W^2/\tau_0)$$

$$q(\psi) = \left[kD_S + D_L + k(D_S - D_L)\psi\right]/2D_L$$

In the equations, the term $\omega(1+\phi_i)/2 \cdot \sum[(1+\phi_j)/2]^2$ in equation (4) represents the interaction between the solid phases at the GBs [35]. $q(\psi)$ in equation (5) is an interpolation function determining the varied diffusion coefficient across the domain.

Ignoring the influence of kinetic undercooling, the calculation parameters of the PF equations could be linked to the physical qualities by expressions: $W = d_0\lambda/a_1$ and $\tau_0 = a_2\lambda W^2/D_L$, where W and $\tau_0$ represent the interface width and relaxation time, which are the length scale and time scale, respectively. In the expressions, $a_1 = 5\sqrt{2}/8$ and $a_2 = 47/75$, $\lambda$ is the coupling constant, $d_0 = \Gamma/|m|(1-k)(c_\infty/k)$ is the chemical capillary length. $\Gamma = \gamma_{sl}T_f/(\rho_s L_f)$ is the Gibbs-Thomson coefficient, where $\gamma_{sl}$ is surface energy between solid and liquid, $T_f$ is the melting point of pure solvent and $L_f$ is the latent heat, respectively.

## 2.2. The macro-micro coupling method

Due to the space-time variational G and R, the microstructure evolution in the melt pool differs with the directional solidification. To predict the microstructures accurately, the variational thermal parameters need to be represented. In current model, the following assumptions are made:

(1) The temperature field is undisturbed by the movement of S/L interface, i.e., the so-called "frozen temperature approximation", which is essentially a statement concerning the relative magnitudes of the terms in the Stefan condition, $\rho_s L_f v_n^* \ll k_{s,l}\nabla T_{s,l}\cdot \mathbf{n}$ [38].

(2) There is no flow in the liquid, which is consistent with the assumption that the densities of the solid and liquid are equal [38].

Based on these assumptions, we adopt two different methods to connect the microscopic PF equations with the macroscopic thermal simulation.

**2.2.1. Thermal parameters G and R (GR model)**

According to literatures [23,31,32], melt pool solidification can be regarded as one kind of directional solidification, whose most important thermal parameters are G and R. Hence we use the G and R to connect the PF equations with the macroscopic thermal process. Since G and R in the pool vary with space and time, the formulations of them should also be space-dependent and time-dependent.

Firstly, equation (4) need to be modified with G and R for representing directional solidification [23]; after that, the steady-state conditions should be replaced by the transient conditions for representing melt pool solidification [39]:

$$a_s^2(\hat{n})\left[1-(1-k)\frac{z-z_0-\int R(t)dt}{l_T(t)}\right]\frac{\partial \phi_i}{\partial t} = \\ \nabla \cdot \left[a_s^2(\hat{n})\vec{\nabla}\phi_i\right] - \partial_x\left(a_s(\hat{n})\cdot a_s'(\hat{n})\cdot \partial_y\phi_i\right) + \partial_y\left(a_s(\hat{n})\cdot a_s'(\hat{n})\cdot \partial_x\phi_i\right) \\ + \phi_i(1-\phi_i^2) - \lambda(1-\phi_i^2)^2\left[U+\frac{z-z_0-\int R(t)dt}{l_T(t)}\right] - \omega\frac{(1+\phi_i)}{2}\sum_{j\neq i}\left(\frac{1+\phi_i}{2}\right)^2 \quad (6)$$

where $l_T(t) = |m|(1-k)(c_\infty/k)/G(t)$ is the thermal length, $m$ is the slope of liquidus line in the phase diagram.

Secondly, the variational G and R should be calculated from the macroscopic thermal simulation. Here is a brief introduction of the calculating method from reference [32]. For GTAW, the shape of melt pool at the longitudinal section can be regarded as the combination of two quarter ellipses, as shown in Figure 1.

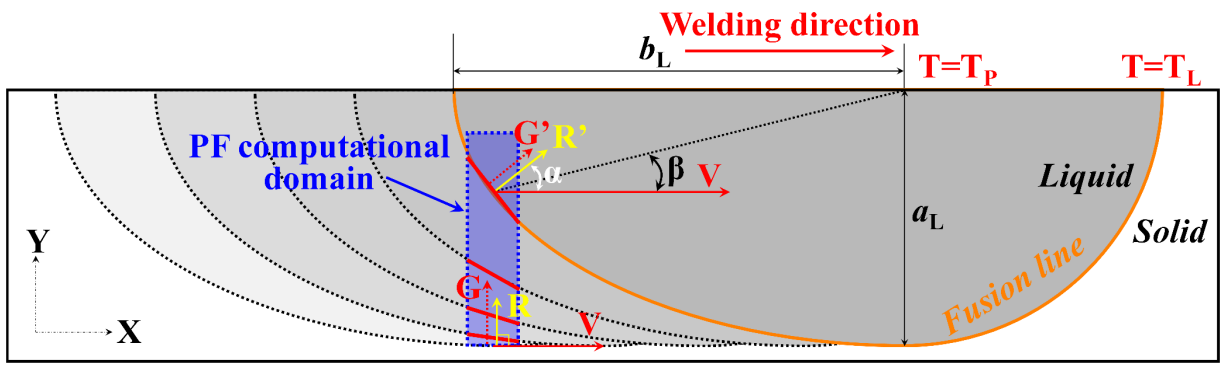

Figure 1. The sketch of the melt pool moving [32]

Figure 1 shows the weld direction is from left to right, so is the movement of pool. The gray regions represent the melt pool, where $a_L$ and $b_L$ are the depth and rear length, respectively. $T_P$ and $T_L$ are the highest temperature at pool center and the liquidus temperature of welded material, respectively. V is the weld speed, α is the intersection angle between R and V, and β is the intersection angle between the radius to the pool center and the x-axis. Assuming the temperature field decreases linearly from pool center to fusion line, the evolutions of G and R can be expressed as [32]:

$$G(\mathbf{r},t) = \frac{T_P - T_L}{\sqrt{x^2(\mathbf{r},t) + y^2(\mathbf{r},t)}} / \cos[\alpha(\mathbf{r},t) - \beta(\mathbf{r},t)] \tag{7}$$

$$R(\mathbf{r},t) = V \cdot \cos\alpha(\mathbf{r},t) \tag{8}$$

where,

$$x(\mathbf{r},t) = V \cdot t - \int V \cdot \cos^2\alpha(\mathbf{r},t) dt \tag{9}$$

$$y(\mathbf{r},t) = \frac{a_L}{b_L} \cdot \sqrt{b_L^2 - x^2(\mathbf{r},t)} \tag{10}$$

$$\alpha(\mathbf{r},t) = \arctan\frac{b_L^2 \cdot y(\mathbf{r},t)}{a_L^2 \cdot x(\mathbf{r},t)} \tag{11}$$

$$\beta(\mathbf{r},t) = \arctan\frac{y(\mathbf{r},t)}{x(\mathbf{r},t)} \tag{12}$$

In the equations, the $a_L$, $b_L$ and $T_P$ could be obtained from the macroscopic thermal simulation, and $T_L$ is related to the welded material. The location parameters x, y, α and β vary with space and time.

By solving equations (7)-(12), we can obtain the G and R at any location of melt pool throughout the entire solidification process of weld.

**2.2.2. Temperature field T (TF model)**

According to literatures [15,16], we can directly adopt the temperature field to connect the PF equations with the macroscopic thermal simulation.

Firstly, the constant temperature field T in equation (4) should be modified to the space-time variational field T(**r**, t) for representing melt pool solidification:

$$a_s^2(\hat{n})\left\{1-\frac{k[T(\boldsymbol{r},t)-T_0]}{|m|c_\infty}\right\}\frac{\partial \phi_i}{\partial t}=$$
$$\nabla\cdot\left[a_s^2(\hat{n})\vec{\nabla}\phi_i\right]-\partial_x\left(a_s(\hat{n})\cdot a_s^{'}(\hat{n})\cdot\partial_y\phi_i\right)+\partial_y\left(a_s(\hat{n})\cdot a_s^{'}(\hat{n})\cdot\partial_x\phi_i\right) \quad (13)$$
$$+\phi_i(1-\phi_i^2)-\lambda(1-\phi_i^2)^2\left\{U+\frac{k[T(\boldsymbol{r},t)-T_0]}{|m|c_\infty(1-k)}\right\}-\omega\frac{(1+\phi_i)}{2}\sum_{j\neq i}\left(\frac{1+\phi_i}{2}\right)^2$$

Secondly, the space-time variational temperature field T(**r**, t) should be calculated from the macroscopic thermal simulation, which can be expressed via an analytical equation [15]. Assuming the shape of melt pool matches the elliptic equations, the temperature field at a given time could be represented as Figure 2.

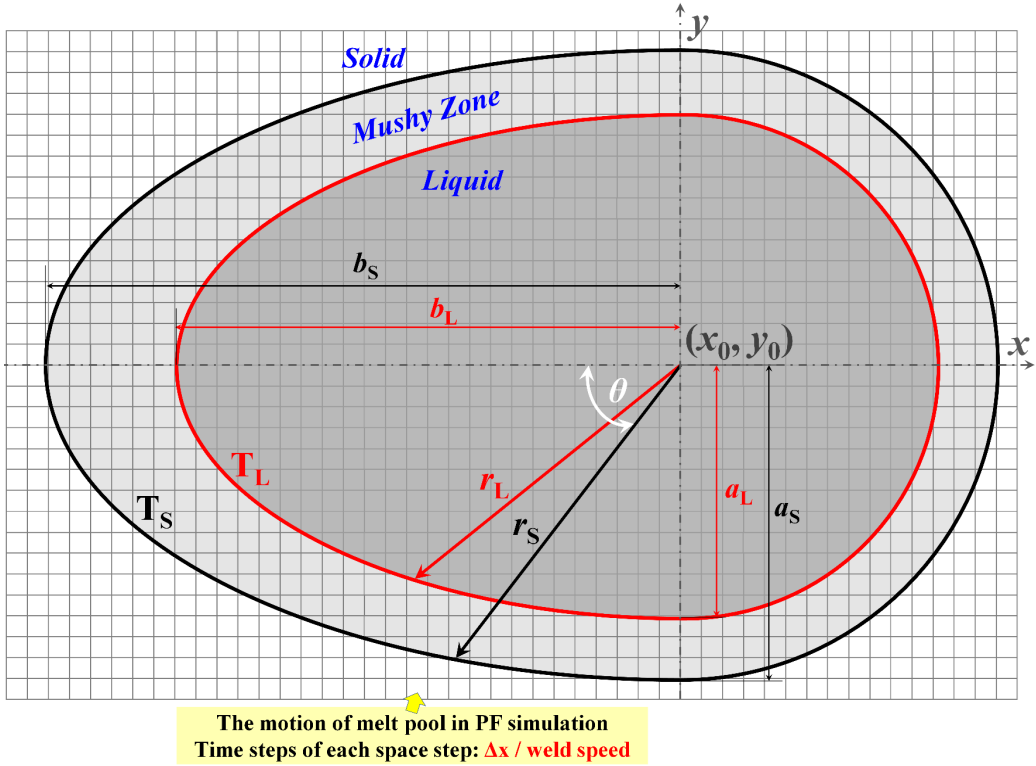

Figure 2. The sketch of elliptic temperature field at one given time

In alloy solidification, the S/L interface appears between liquidus isothermal line ($T_L$) and solidus line ($T_S$), hence we determine the temperature field based on the $T_L$ and $T_S$. Simplifying the temperature field decreases from pool center to fusion line, the expression of temperature field can be given by:

$$T(\boldsymbol{r},t)\overset{\text{at a given time}}{\equiv}T(x,y)=T_L+(T_S-T_L)\cdot\frac{r(x,y)-r_L}{r_S-r_L} \quad (14)$$

The distance from ellipse center to one point in the pool is:

$$r(x,y)=\sqrt{(x-x_0)^2+(y-y_0)^2} \quad (15)$$

According the expression of elliptic equations, the distance from pool center to one point at the liquidus isothermal line is:

$$r_L = \sqrt{\frac{a_L^2 \cdot b_L^2}{a_L^2 \cdot \sin^2 \theta + b_L^2 \cdot \cos^2 \theta}} \quad (16)$$

Similarly, the distance from pool center to one point at the solidus line is:

$$r_S = \sqrt{\frac{a_S^2 \cdot b_S^2}{a_S^2 \cdot \sin^2 \theta + b_S^2 \cdot \cos^2 \theta}} \quad (17)$$

By solving equations (14)-(17), we can obtain the distribution of temperature across the pool. Assuming the movement of melt pool does not cause the temperature field fluctuating with time, we could obtain the temperature field at any location of melt pool throughout the entire solidification process of weld.

### 2.3. Computational procedure

#### 2.3.1. Macroscopic simulation of the thermal process

In the thermal simulation, the dimensions of the welded plate are set to be 100mm×60mm×2mm. The welded material is Al-2.0wt.%Cu and the process is GTAW without filler metal. The welding parameters are: current 100A, arc voltage 12V and weld speed 1.0mm/s.

The thermal process was simulated through the commercial FE software MSC.MARC®. The detailed simulation procedure can be found in reference [15]. Based on the macroscopic simulation, we could obtain the thermal parameters for the microscopic PF simulations.

#### 2.3.2. Microscopic simulation of solidification

In the PF simulation, the material Al-2.0wt.%Cu could be regarded as a dilute binary alloy, whose material parameters are shown in Table 1 [30,40,41].

As for the calculation parameters, the most important parameter in PF simulation is the interface width W [21,42]. Specifically, the accuracy of simulation increases with the decrease of W, while the computational cost dramatically increases with the decrease of W. According to literature [23,43], W just needs to be one order of magnitude smaller than the characteristic length scale of microstructures. The characteristic length for alloy solidification is $L_C \sim \sqrt{d_0 \ast D_L/v_{tip}}$ [38], hence W is set to be 0.18μm in this paper. The Neumann boundary conditions with zero-flux were used for both the phase field and supersaturation field. The time step size was chosen below the threshold of numerical instability for diffusion equation, i.e., $\Delta t < (\Delta x)^2/(4D_L)$. The current study used fixed grid size $\Delta x = 0.8W$ and time step size $\Delta t = 0.01\tau_0$.

Table 1. The material parameters of Al-2.0wt.%Cu for the PF simulation [30,40,41]

| Symbol | Value | Unit |
|---|---|---|
| Liquidus temperature, $T_L$ | 927.8 | K |
| Solidus temperature, $T_S$ | 896.8 | K |
| Diffusion coefficient in liquid phase, $D_L$ | $3.0\times10^{-9}$ | m²/s |
| Diffusion coefficient in solid phase, $D_S$ | $3.0\times10^{-13}$ | m²/s |
| Alloy composition, $c_\infty$ | 2.0 | wt.% |
| Liquidus slope, m | –2.6 | K/wt.% |
| Gibbs-Thomson coefficient, Γ | $2.4\times10^{-7}$ | K·m |
| Anisotropy of surface tension, $\varepsilon_4$ | 0.010 | |
| Equilibrium partition coefficient, k | 0.14 | |
| Grain interaction parameter, ω | $8.0*(T-T_S)/(T_L-T_S)$ | |

Moreover, to consider the infinitesimal perturbation of thermal noise on the S/L interface, a fluctuating current $J_U$ is introduced on the right-hand side of the diffusion equation. By using the Euler explicit time scheme, we have [44,45]:

$$U^{t+\Delta t} = U^t + \Delta t \left( \partial_t U - \vec{\nabla} \cdot \vec{J}_U \right) \qquad (18)$$

The components of $J_U$ are random variables obeying a Gaussian distribution, which has the maximum entropy relative to other probability distributions [46]:

$$\left\langle J_U^m(\vec{r},\vec{t}) J_U^n(\vec{r}\,',\vec{t}\,') \right\rangle = 2D_L q(\psi) F_U^0 \delta_{mn} \delta(\vec{r}-\vec{r}\,') \delta(t-t') \qquad (19)$$

During the numerical simulation, the discretized noise in 2D becomes [44,45]:

$$\vec{\nabla} \cdot \vec{J}_U \approx \left( J_{x,i+1,j}^n - J_{x,i,j}^n + J_{y,i,j+1}^n - J_{y,i,j}^n \right) / \Delta x \qquad (20)$$

In addition, the constant noise magnitude $F_u^0$ is defined as [45]:

$$F_U^0 = \frac{k v_0}{(1-k)^2 N_A c_\infty} \qquad (21)$$

$F_U^0$ is the value of $F_U$ for a reference planar interface at temperature $T_0$, where $v_0$ is molar volume of the solute atoms, and $N_A$ is the *Avogadro* constant.

### 2.3.3. Multi-scale simulation of melt pool solidification

The program code of the multi-scale simulation was written by C++ and executed on the platform of π 2.0 cluster, supported by the Center for High Performance Computing at the Shanghai Jiao Tong University (SJTU). The macroscopic thermal parameters were discretized to the meshes matching the microscopic PF simulation. The governing equations were solved by the explicit Finite Difference Method (FDM), and the Message Passing Interface (MPI) parallelization was adopted to improve the computational efficiency.

## 3. Results and discussion

### 3.1. Single-crystal solidification at a local region of pool

To compare the two models mentioned in section 2.2, the single-crystal solidification at a local region of pool was simulated through the GR model and TF model respectively, for which the PCO is parallel with the y-axis. The computational domain is 2000×5000 grids, corresponding to 288.0μm×720.0μm in real unit. It takes about 16 hours using 40 cores to finish one job.

The evolutions of single-crystal solidification under the space-time variational conditions are shown in Figure 3, which is located at the bottom of melt pool.

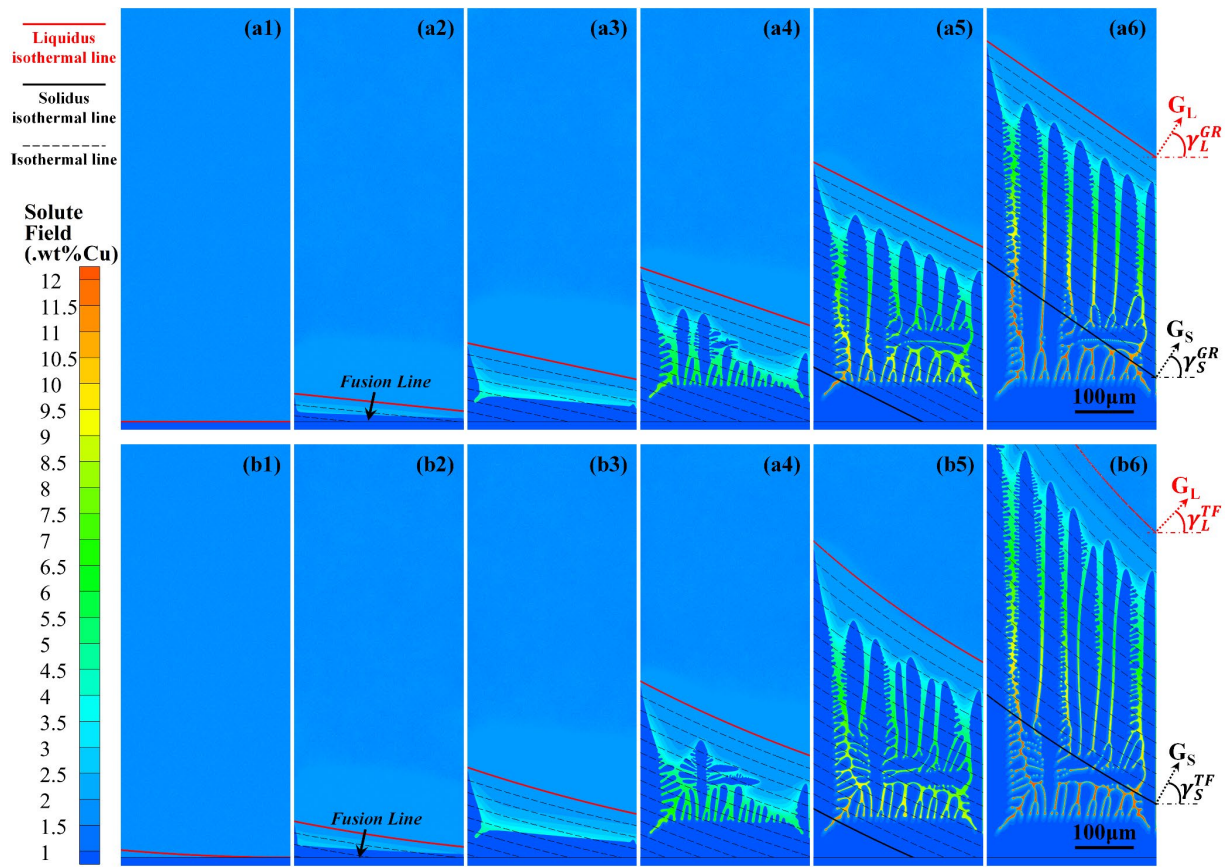

Figure 3. The single crystal solidification at local region under the space-time variational conditions: (a) GR model; (b) TF model ($t_1$ = 0.0s; $t_2$ = 0.344s; $t_3$ = 0.688s; $t_4$ = 1.032s; $t_5$ = 1.376s; $t_6$ = 1.677s.)

In Figure 3, the red lines are liquidus thermal lines $T_L$, the black solid lines are solidus isothermal lines $T_S$, the black dash lines are the isothermal lines between $T_L$ and $T_S$, and the temperature difference between each two lines is 2K. The distance between the front interface and $T_L$ indicates the front position of interface departs from the position of $T_L$, i.e., the tip undercooling, representing the non-equilibrium of solidification. In addition, the heat source moving makes the thermal parameters space-time variational. As shown in Figure 3, the pool moves from left to right, the isothermal lines also deflect to right. As a result, the solidification evolution shows deflection characteristics, including the morphological evolution of interface and the degree of solute segregation. These special phenomena will be discussed in the following.

**3.1.1. The necessity of considering space-time variational conditions**

As shown in Figure 3, both the GR model and TF model represent the space-time variational conditions, including the magnitude and direction of G and R. Specifically, the distance between each two isothermal lines reflects the magnitude of G. The increasing distance between the lines means the temperature gradient G decreases with time. The moving speed of the isothermal line reflects the magnitude of R. The increasing moving speed of the lines means the pulling speed R increases with time. Meanwhile, the normal directions of the isothermal lines represent the directions of G and R, from vertical to right. In conclusion, by deflecting the isothermal lines, both the GR model and TF model could represent the variational G and R in the pool.

Due to the space-time variational G and R, the microstructure evolution shows specific characteristics. At the beginning, the S/L interface keeps planar and advances slowly to the liquid, as shown in Figure 3(a1)-(a2) and (b1)-(b2). As time goes on, the solute aggregation at the S/L interface makes the planar instability occur, i.e., the planar transform to the cellular, shown in Figure 3(a3)-(a4) and (b3)-(b4). The variational G and R make the cellular grow gradually from the tail to the bottom rather than grow periodically along the x-axis. That is, the microstructure evolution in the melt pool is non-periodic, different from the PF simulations without considering the isothermal lines deflecting [31,47,48]. Moreover, the non-periodic characteristics of the evolution are increasingly obvious with time, including the deflections of the morphologies and solute concentration, shown in Figure 3(a5)-(a6) and (b5)-(b6). The non-periodic phenomena indicate the necessity of considering the isothermal lines deflecting, i.e., the space-time variational solidification conditions in the pool. Meanwhile, both the TF model and GR model could represent the variational conditions.

### 3.1.2. The differences between the GR model and TF model

Even they both could represent the space-time variational conditions in the pool, the GR model and TF model still show differences. Comparing Figure 3(a1)-(a6) with Figure 3(b1)-(b6), the differences between the two models include the thermal parameters (isothermal lines) and microstructure evolution, which result from the different parameters connecting the PF equations with macroscopic thermal simulation. Specifically, the GR model uses the parameters G and R. By solving equations (7)-(12), we can calculate G and R at any location of the pool throughout the entire solidification. However, at a given time, we can only calculate G and R at one specific point of pool at the macroscopic scale. When solving the PF equations, we extrapolate the G and R from one point to the whole computational domain based on the slope of the solidus isothermal line. As a result, the GR model makes the isothermal lines into a set of parallel lines across the domain, as shown in Figure 3(a1)-(a6). By contrast, the TF model uses the temperature field T(**r**,t) instead of G and R. By solving equations (14)-(17), we can obtain the temperature field at any location of the pool throughout the entire solidification. The temperature field is discretized from the macroscopic scale to the microscopic PF meshes. As a result, the TF model makes the isothermal lines into a set of concentric elliptic lines, shown in Figure 3(b1)-(b6). In conclusion, the GR model uses a set of parallel lines to reflect the isothermal lines, while the TF model uses a set of concentric elliptic lines to reflect the isothermal lines.

To reflect the distinctions between the two models intuitively, the temperature field in current study was represented in Figure 4(a), as well as the sketch of the connecting between the PF simulation and the thermal parameters. It is obvious it is too simplified to regard all the isothermal lines as a set of parallel lines in the GR model, especially at the late stage. To compare the isothermal lines clearly, we capture the evolutions of intersection angles between the normal directions of isothermal lines and the x-axis. The definitions of the angles are sketched in Figure 3(a6) and (b6), which are solidus isothermal line angle $\gamma_S^{GR}$ and liquidus isothermal line angle $\gamma_L^{GR}$ from the GR model, solidus isothermal line angle $\gamma_S^{TF}$ and liquidus isothermal line angle $\gamma_L^{TF}$ from the TF model, respectively. The evolutions of the angles are shown in Figure 4(b), at the very beginning, the angles are almost 90°, indicating the isothermal lines are almost horizontal. Then the angles decreases, representing all the lines deflect to the right side as time goes on. Specifically, the decrease rates of $\gamma_S^{GR}$ and $\gamma_L^{GR}$ keep the same all the time, due to the parallel isothermal lines, while the decrease rates of $\gamma_S^{TF}$ and $\gamma_L^{TF}$ differ with each other as time goes on, due to the concentric elliptic isothermal lines. Moreover, at the beginning, all the intersection angles are almost the same, meaning the differences between the GR model and TF model are ignorable. As time goes on, $\gamma_S^{TF}$ decreases slower than $\gamma_S^{GR}$, which reflects the abstract values of the slope of $\gamma_S^{TF}$ are smaller than those of $\gamma_S^{GR}$. By contrast, $\gamma_L^{TF}$ decreases faster than $\gamma_L^{GR}$, which reflects the abstract values of the slope of $\gamma_L^{TF}$ are larger than those of $\gamma_L^{GR}$. The different evolutions of intersection angles directly represent the different evolutions of isothermal lines. Correspondingly, the microstructures in the two models are also different, shown in Figure 4(c1)-(c6) and (d1)-(d6), including the morphological evolution and the degree of solute segregation. Due to the faster deflecting speed of liquidus line in the TF model, the dendrites in the TF model grow faster than the GR model. The solute concentration also follows the isothermal lines deflecting, showing non-periodicity and deflection, shown in Figure 4(c1)-(c6) and (d1)-(d6). The detailed analysis of the tip data will be carried out in the future.

In conclusion, for single-crystal solidification at the local region of pool, the non-periodic evolutions indicate the necessity of considering the variational conditions, meanwhile both the GR model and TF model can reflect the variabilities. On the other hand, the TF model can represent the variational conditions better than the GR model.

It should be noted the PF simulation of solidification at the local region may be affected by the boundary effect, illustrated by the evolution of the leftmost dendrite A in Figure 4(c) and (d).

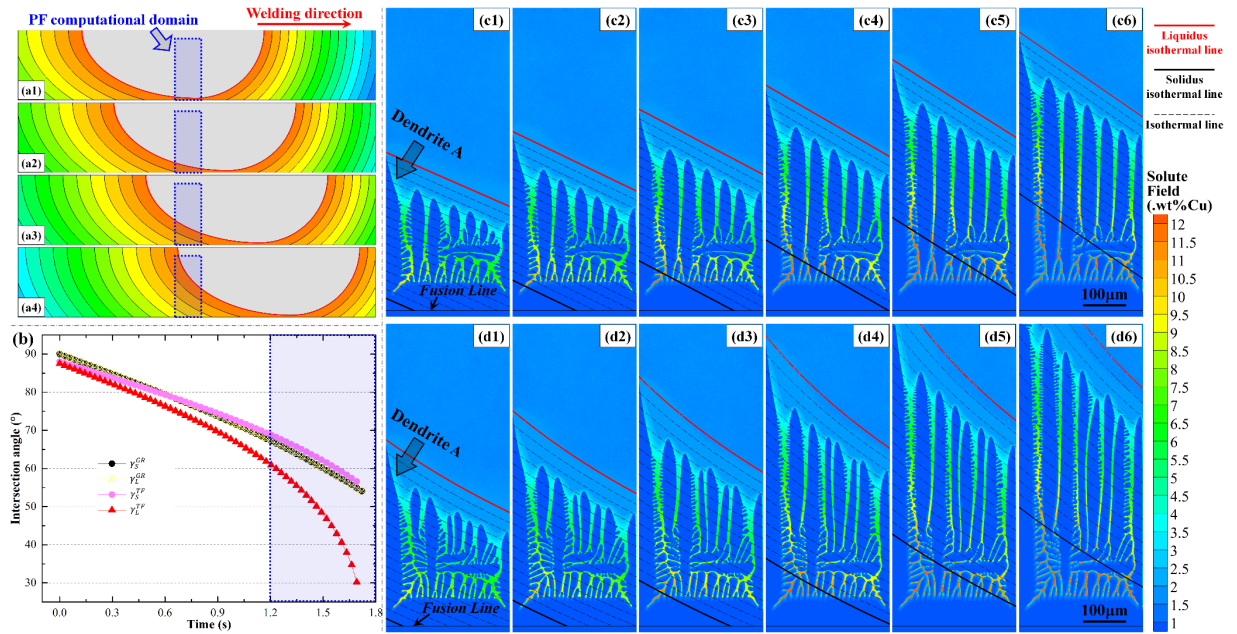

Figure 4. (a) The sketch of the connecting between the microscopic PF simulation and temperature field; (b) The evolutions of intersection angles between the normal directions of isothermal lines and x-axis; (c) The single-crystal solidification from the GR model; (d) The single-crystal solidification from the TF model ($t_1$ = 1.247s; $t_2$ = 1.333s; $t_3$ = 1.419s; $t_4$ = 1.505s; $t_5$ = 1.591s; $t_6$ = 1.677s.).

## 3.2. Single-crystal solidification at the whole region of pool

To avoid the boundary effect, the pool shape is set to be an ellipse, different from the reference [11]. The thermal parameters in the pool for the PF simulation are shown in Table 2. To investigate the influence mechanism of the crystallographic parameters (the interface energy anisotropy and PCO) and solidification parameters (G and R) on the microstructures, we simulate the solidification processes of single-crystal with different PCOs. Due to the four-fold symmetry of cubic crystal, the PCOs are set to be 0°, 15°, 30° and 45°, respectively. The computational domain is 9600×5100 grids, corresponding to 1382.4μm×734.4μm in the real unit. It takes about 33 hours using 80 cores to finish one job.

The evolutions of single-crystal solidification at the whole region of melt pool are shown in Figure 5, including the temperature field and solidification structures with different PCOs.

Figure 5(a) represent the evolution of temperature field in the pool, where the black solid lines are the isothermal lines. The melt pool moves from left to right, hence the dendrites also grow from left to right, shown in Figure 5(b)-(e), where the red solid lines are $T_L$, the black solid lines are $T_S$. The black dash lines are the isothermal lines between $T_L$ and $T_S$, and the temperature difference between each two lines is 2K. The distance between the S/L interface and $T_L$ represents the tip undercooling, indicating the non-equilibrium of melt pool solidification. On the other hand, by comparing Figure 5(b)-(e), the evolutions of the single-crystal solidification differ with each other, resulting from the different PCOs. In addition, the space-time variational parameters across the domain make the solidification at the whole region differ with the solidification at the local region in section 3.1.

The space-time variabilities of the solidification parameters across the pool, as well as the evolutions of the single-crystal with different PCOs, will be discussed in the following.

Table 2. The thermal parameters of melt pool for the PF simulation

| Symbol | Value | Unit |
| --- | --- | --- |
| Depth of liquidus isothermal line, $a_L$ | 0.35 | mm |
| Depth of solidus isothermal line, $a_S$ | 0.60 | mm |
| Forward length of liquidus isothermal line, $b_L$ | 0.70 | mm |
| Forward length of solidus isothermal line, $b_S$ | 1.20 | mm |
| Welding speed | 1.00 | mm/s |

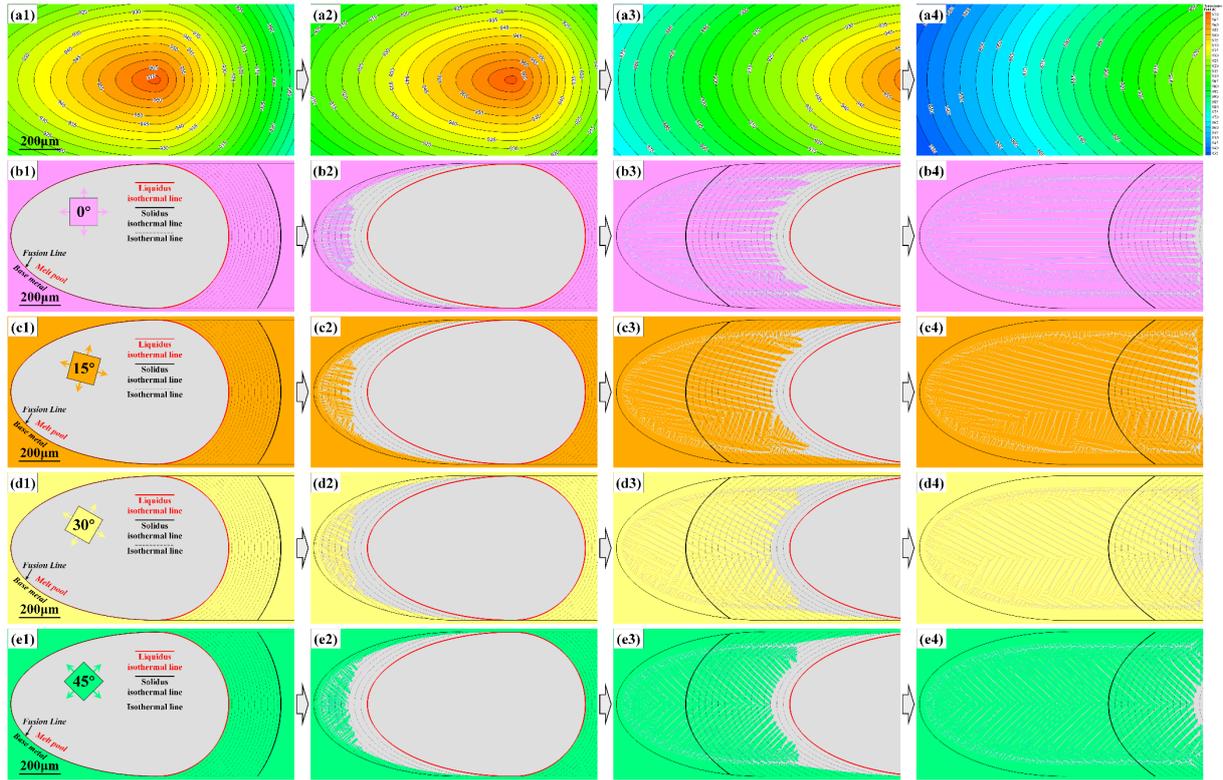

Figure 5. The single-crystal solidification at whole region of pool: (a) The evolution of temperature field in the pool; (b) The morphological evolution of single-crystal (0°); (c) The morphological evolution of single-crystal (15°); (d) The morphological evolution of single-crystal (30°); (e) The morphological evolution of single-crystal (45°) ($t_1$ =0.0s; $t_2$ = 0.258s; $t_3$ = 0.831s; $t_4$ = 1.405s).

### 3.2.1. The space-time variabilities of solidification parameters

We obtained the space-time distributions of G and R in the pool. As representative, the distributions of G and R at the given time (t = 0.258s) are shown in Figure 6. The temperature gradient G decreases gradually from bottom to tail of the pool, while the pulling speed R increases gradually from bottom to tail of the pool. The arrow directions represent the directions of G and R, along the normal directions of the isothermal lines, i.e., the Thermal Gradient Direction (TGD). Due to the distribution of temperature field in the pool, the TGDs vary with the location, and all the TGDs point to the pool center. On the other hand, the parameters G and R dominate the microstructure evolution, shown in Figure 7(a) [2]. The multiplication G*R determines the size of microstructure, while the ratio G/R determines the morphology of microstructure, i.e., the planar, cellular, columnar dendritic and equiaxed dendritic. We calculate the distributions of G*R and G/R, shown in Figure 7(b) and (c). The multiplication G*R increases gradually from bottom to tail of the pool, while the ratio G/R decreases gradually from the bottom to the tail. In conclusion, at the given time, the parameters G, R, G*R and G/R vary continuously across the domain, indicating the space-variability of solidification conditions in the melt pool.

As for the time-variability, we chose the characteristic points along fusion line, from the bottom to the tail, and obtain the evolutions of solidification parameters at these points. As shown in Figure 8(a), points A to H locate along the fusion line and the intersection angle between each two lines is 15°. Then the evolutions of parameters G, R, G*R and G/R at the chosen points are obtained, as shown in Figure 8(b)-(e). Taking point A as an example, G decreases with time while R increases with time, G*R increases with time while G/R decreases with time. On the other hand, at one given time, G decreases from point A to point H (from the bottom to the tail) while R increases from the bottom to the tail, G*R increases from the bottom to the tail while G/R decreases from the bottom to the tail. The evolutions of G, R, G*R and G/R at the different chosen points, shown in Figure 8, indicate the time-variability and space-variability of solidification parameters in the pool. Moreover, the variation rates of the parameters at points A to H differ with each other. Specifically, the parameters at point H (the tail of pool) remain stable all the time while the parameters at point A (the bottom of pool) change considerably with time. As time goes on, the parameters at different points tend to be similar, shown by the convergent curves in Figure 8(b)-(e).

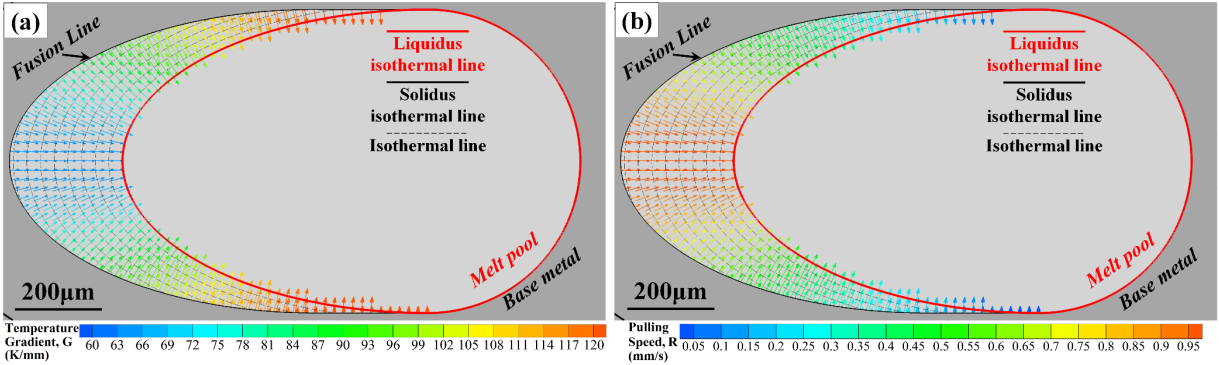

Figure 6. The distributions of solidification parameters (a) temperature gradient T and (b) pulling speed R across the melt pool at a given time (t = 0.258s).

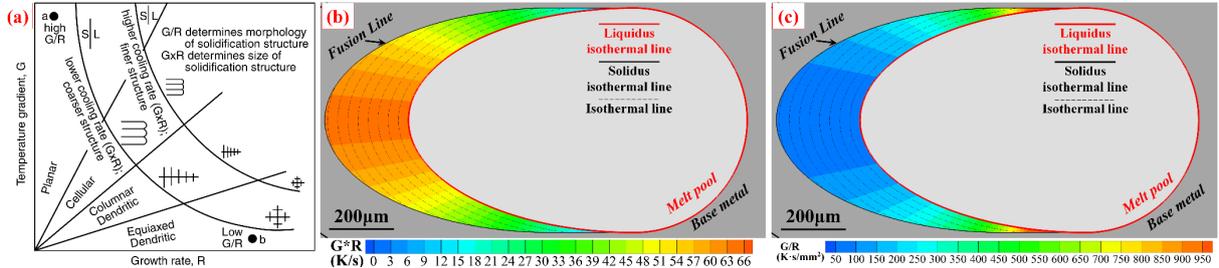

Figure 7. (a) The effect of solidification parameters G and R on the microstructures [2]; (b) the distribution of the ratio G/R across the pool; (c) the distribution of the multiplication G*R at a given time (t = 0.258s).

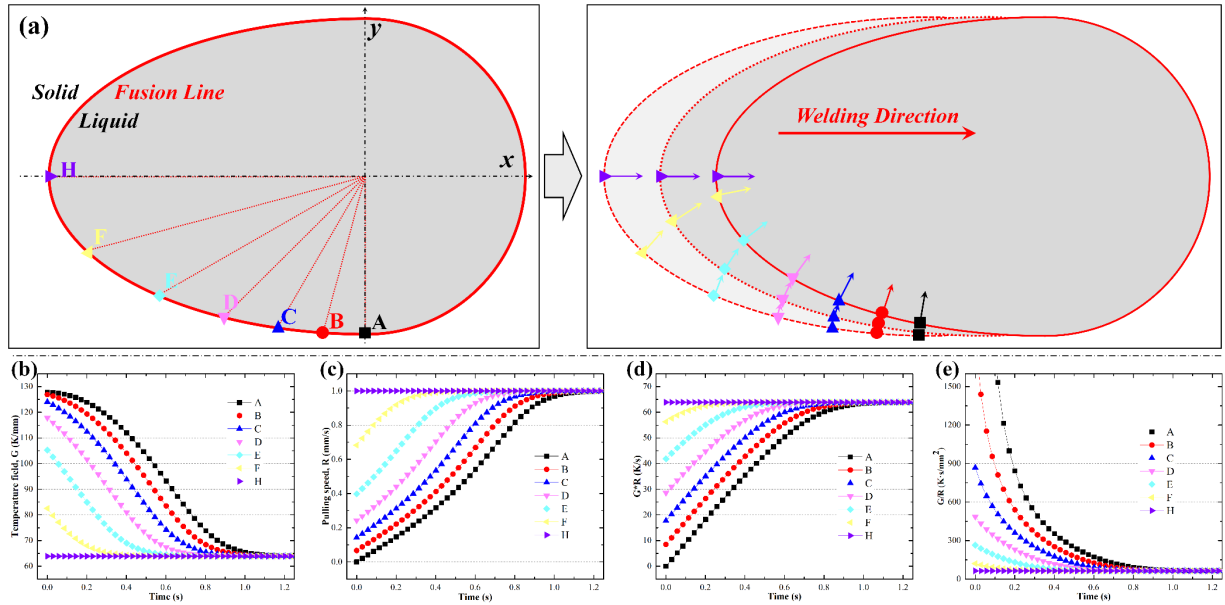

Figure 8. The evolutions of solidification parameters at different points: (a) the chosen points A to H along fusion line; (b) the evolution of thermal gradient G; (c) the evolution of pulling speed R; (d) the evolution of G*R; (e) the evolution of G/R.

In conclusion, the evolutions of parameters G, R, G*R and G/R demonstrate the space-time variabilities of conditions across the pool. Moreover, the parameters at the tail of pool remain stable while the parameters at the bottom of pool are unstable. As time goes on, the parameters at different locations tend to be similar.

**3.2.2. Single-crystal solidification under variational conditions**

As for microstructure evolution, according to Figure 7(a), we can know the possible solidification modes across the pool, where the microstructures go through the planar-cellular-columnar-equiaxed transition [2]. It should be noted the equiaxed grains only appear under the relatively low G and high R [2]. The weld speed in current study is 1.0mm/s, corresponding to low R, hence the CET is ignored in this paper. We focus on the planar-cellular-columnar transition, i.e., the epitaxial growth and competitive growth.

**(1) The epitaxial growth**

At the beginning of solidification in a weld without filler metal, along the fusion line, nucleation occurs by arranging atoms from the liquid metal upon the substrate grains without altering their existing PCOs, i.e., the epitaxial growth [2]. The single-crystal solidification in epitaxial growth is shown in Figure 9, where the PCO is parallel with the x-axis. The pool moves from left to right, the red solid lines are $T_L$, the black dash lines are the isothermal lines between $T_L$ and $T_S$, and the temperature difference between each two lines is 2K. Due to the space-time variational parameters across the domain, the evolution at the whole region differ with that at the local region (in section 3.1.).

The solidification mode is dominated by the G*R and G/R, we obtain the morphological evolution and the distributions of G*R and G/R, in Figure 9(a) and (b). At the beginning, the S/L interface keeps planar and advances slowly to liquid. As time goes on, the accumulation of solutes makes the planar instability appear, represented by the transition from the planar to the cellular. As shown in Figure 9(a), the G/R decreases from the tail to the bottom, while the planar instability at the tail appears earlier than the bottom. That is, the smaller the G/R is, the earlier the planar instability occurs, consistent with Figure 7(a) and literature [49].

In the other hand, the solute aggregation ahead of interface makes the planar instability appear, where solute aggregation results from solute trapping, determined by the degree of non-equilibrium. Thus, we move on to the degree of non-equilibrium in the pool. As shown in Figure 9(b), the G*R, i.e., the cooling rate (unit: K/s), decreases from the tail to the bottom. As a result, the degree of non-equilibrium decreases from the tail to the bottom, shown by the distribution of solute field in Figure 9(c). Meanwhile, higher degree of solute segregation make the excess Gibbs energy at the interface decrease, corresponding to lower interface energy [50]. With the influences of interface energy and its anisotropy, the planar instability occurs. Higher degree of solute segregation corresponds to earlier onset of planar instability, in Figure 9, the planar instability appears at the tail firstly and then expands to other locations along the interface. That is, the solute segregation and interface anisotropy are the main factors of instability.

To test this conclusion, under the same weld conditions, the single-crystal solidification with different PCOs are carried out, shown in Figure 10. At the planar growth stage, in Figure 10(a1), (b1), (c1), and (d1), the interface energy and its anisotropy do not affect the solute diffusion and the pulling distance of interface, which is consistent with literature [51]. That is, the PCO of grain has little influence on the evolution at this stage. Meanwhile, due to the variational parameters, the cooling rate (G*R) increases from the bottom to the tail, determining the degree of non-equilibrium and the solute segregation. The solute segregation at the tail is the highest, then it decreases from the tail to the bottom. Higher solute segregation reduces the excess free energy, decreasing the interface energy and making the planar instability occur. Hence, the planar instability appears at the tail firstly and then expands to other locations along the interface, illustrated in Figure 10(a2), (b2), (c2), and (d2), showing few differences between the simulations with different PCOs. The results verify the discussion that solute segregation influences the interface energy and results in the interface instability. The similarity in the simulations with different PCOs indicates, at the planar growth stage, the influence of solidification parameters is more significant than crystallographic parameters, consistent with literature [49].

In conclusion, at the epitaxial growth stage, the planar instability is dominated by solute segregation, through reducing the excess free energy at the S/L interface and decreasing the interface energy. When the interface energy reduces to a critical level, the planar instability occurs. On the other hand, at this stage, the effect of solidification parameters is more significant than crystallographic parameters.

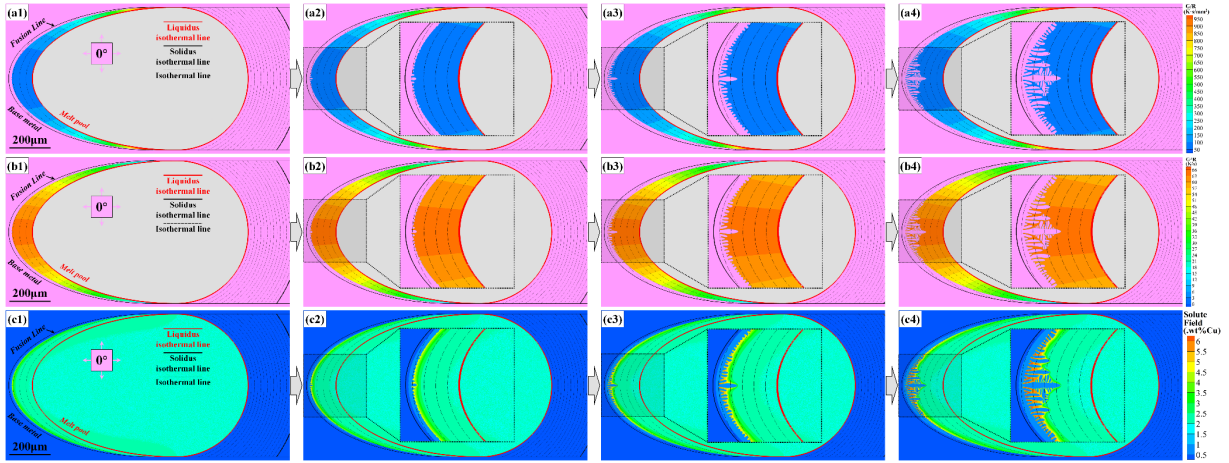

Figure 9. The single-crystal solidification at the epitaxial growth stage: (a) the morphological evolution under the variational G*R; (b) the morphological evolution under the variational G/R; (c) the evolution of solute field. ($t_1$ = 0.115s; $t_2$ = 0.143s; $t_3$ = 0.172s; $t_4$ = 0.201s.)

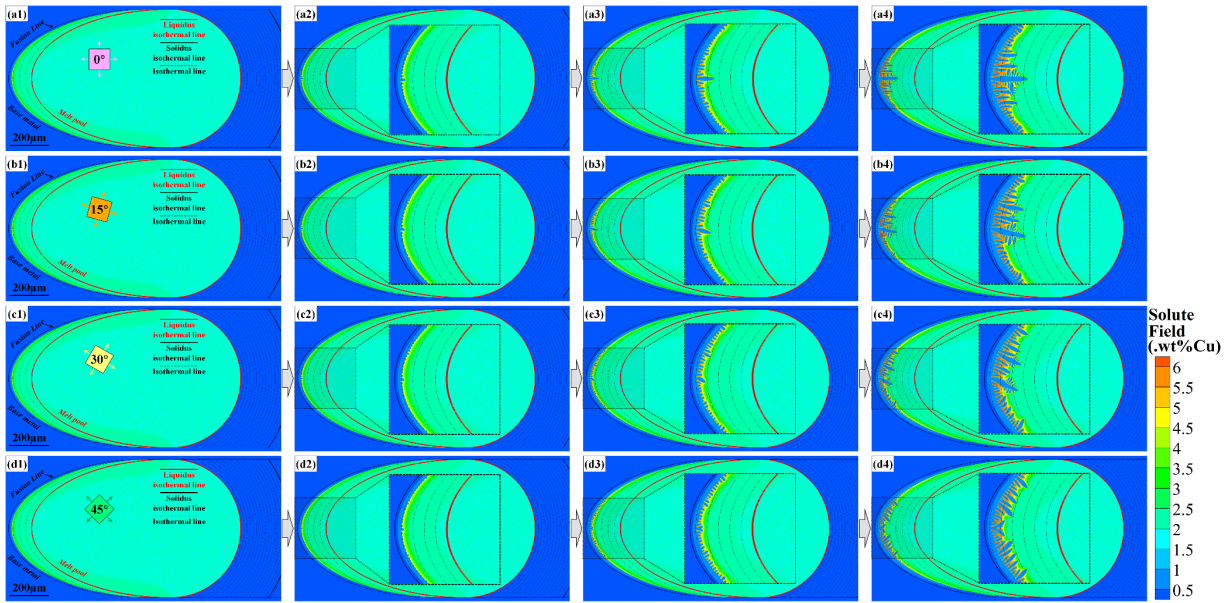

Figure 10. The evolutions at the epitaxial growth stage: (a) The solute-field evolution of single-crystal (0°); (b) The solute-field evolution of single-crystal (15°); (c) The solute-field evolution of single-crystal (30°); (d) The solute-field evolution of single-crystal (45°). ($t_1$ = 0.115s; $t_2$ = 0.143s; $t_3$ = 0.172s; $t_4$ = 0.201s.)

**(2) The competitive growth**

Although the single-crystal has the same PCO, due to the non-planar interface and the variational G and R in the melt pool, the cellular encounter and collide with each other. The process turns into the competitive growth, shown in Figure 11(a)-(d), whose PCOs are 0°, 15°, 30° and 45°, respectively.

Comparing the simulations in Figure 11, the evolutions show differences, due to the effect of interface energy, which is determined by the PCO and anisotropic strength, shown in equation (3). Specifically, Figure 11(a) shows the [100] dendrites paralleled with the weld direction. Due to the ellipse-shape of pool, the [100] dendrites along the axial direction (parallel with the weld direction) compete with the [010] dendrites along the radial direction (vertical with the weld direction). Under the low weld speed, the [100] dendrites along the axial direction has significant advantages in the competition and survive to the end. On the other hand, due to the stochastic factors in growth, the morphologies are not symmetric with the axial direction. Figure 11(b) shows the single-crystal solidification, the angle between whose [100] direction and the weld direction is 15°. Due to the low weld speed, the dendrites along the radial direction cannot survive in the competition, while the dendrites along the axial direction grow to the end. On the other hand, among the dendrites along the axial direction, they compete with each other. The [010] dendrites, evolving from the sidebraches of some [100] dendrites, compete with other [100] dendrites. The competition makes the fluctuation of interface, shown by red dotted line in Figure 11(b4). The evolution of single-crystal, whose [100] direction is at an angle of 30° with the weld direction, shown in Figure 11(c). Similarly, the dendrites along the axial direction grow to the end and the interface shows fluctuation due to the competition. Figure 11(d) shows the single-crystal, whose [100] direction is 45° with the weld direction. The dendrites from the fusion line are eliminated, the dendrites from the tail grow to the end. Moreover, among the dendrites from the tail region, the new [100] dendrites compete with the new [010] dendrites, and the interface always stays near the centerline.

In conclusion, at the competitive growth stage, the evolutions with different PCOs show differences, due to the effect of interface energy anisotropy. Despite the differences, the common of the simulations is that the dendrites from the fusion line are eliminated, while the dendrites from the tail grow to the end. This phenomenon results from the low weld speed. On the other hand, the pulling distances of the planar crystal shows few differences between the simulations, shown by the black dotted lines in Figure 14(a4), (b4), (c4), and (d4). This phenomenon indicates, at the initial planar growth stage, the effect of solidification parameters on the initial instability is greater than crystallographic parameters.

In should be noted, although the simulation of single-crystal can investigate the effects of solidification parameters and crystallographic parameters separately, the welded materials in practice are poly-crystal rather than single-crystal. The simulation of poly-crystal can reproduce the melt pool solidification more realistically, carried out in the next section.

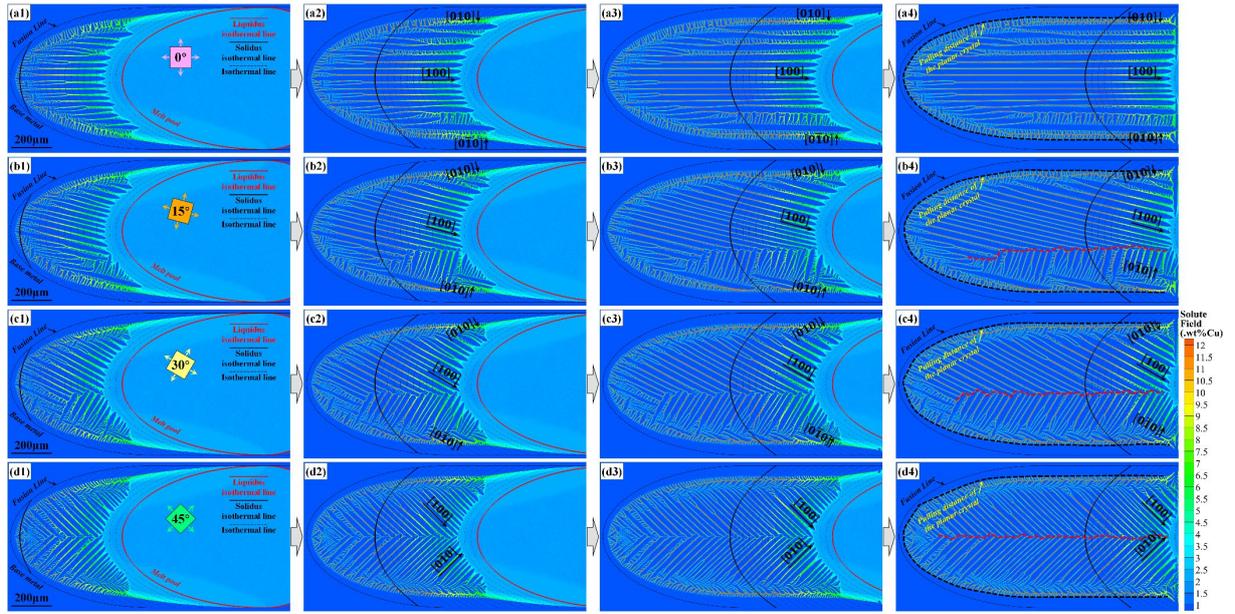

Figure 11. The single-crystal solidification at the competitive growth stage: (a) the evolution of solute field with PCO 0°; (b) the evolution of solute field with PCO 15°; (c) the evolution of solute field with PCO 30°; (d) the evolution of solute field with PCO 45°. ($t_1$ = 0.545s, $t_2$ = 0.831s, $t_3$ = 1.118s, $t_4$ = 1.405s.)

**3.3. Poly-crystal solidification at the whole region of pool**

The thermal parameters for the poly-crystal simulation are the same as the single-crystal simulation, shown in Table 2. The initial grains are set to be random distribution, shown in Figure 12(b), where the colors of grain reflect the intersection angles $\theta_i^0$ between the PCOs and the x-axis. Due to the four-fold symmetry of cubic crystal, $\theta_i^0$ are set to be 0°, 15°, 30°, 45°, 60° and 75°, respectively. The computational domain is 9600×5100 grids, corresponding to 1382.4μm×734.4μm in the real unit. It takes about 62 hours using 80 cores to finish one job.

The poly-crystal solidification at the whole region of pool is shown in Figure 12, including temperature field, grain, and sub-grain. The pool moves from left to right, the grains also grow from left to right. The red solid lines are $T_L$, the black solid lines are $T_S$. The black dash lines are the isothermal lines between $T_L$ and $T_S$, and the temperature difference between each two lines is 2K. The distance between the interface and $T_L$ reflects the tip undercooling and indicates the non-equilibrium of solidification. On the other hand, due to the effect of GBs, the poly-crystal solidification differs with the single-crystal, which will be discussed in the following. Similar with the previous section, the CET is ignored in the poly-crystal solidification. We focus on the planar-cellular-columnar transition, i.e., the epitaxial growth stage and competitive growth stage.

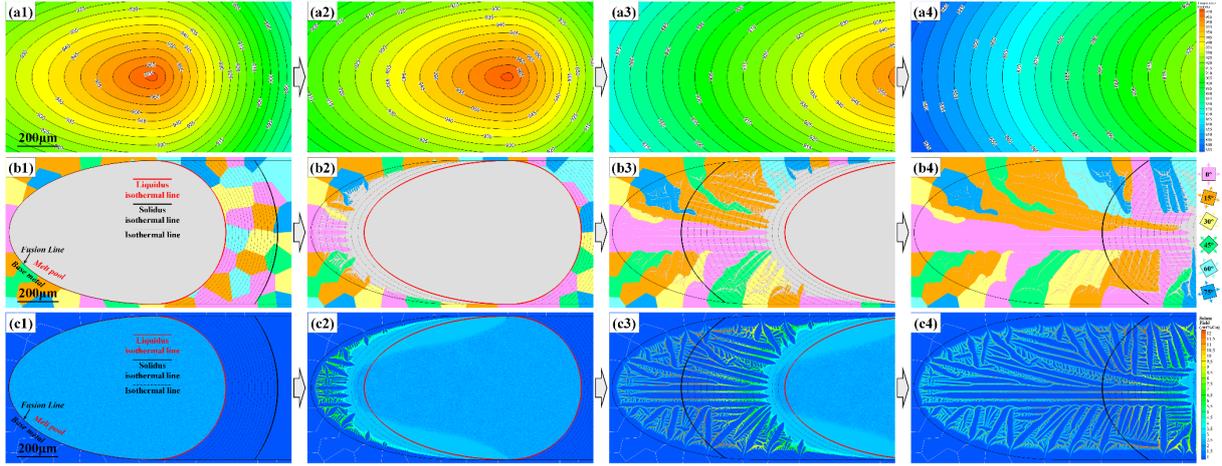

Figure 12. The poly-crystal solidification at whole region of pool: (a) The evolution of temperature field in the pool; (b) The evolution of grain; (c) The evolutions of sub-grain and solute field. ($t_1 = 0.0s$; $t_2 = 0.258s$; $t_3 = 0.831s$; $t_4 = 1.405s$).

### 3.3.1. The epitaxial growth

To investigate the effect of GBs, the evolutions of single-crystal and poly-crystal are shown in Figure 13(a) and (b), respectively. On the one hand, due to the space-time variational G and R, the initial instability appears at the tail firstly, and then expands to other locations, both in the single-crystal solidification and poly-crystal solidification, shown by the red circle-marked regions in Figure 13(a2)-(a3) and (b2)-(b3). On the other hand, due to the effect of GBs in the poly-crystal solidification, the initial instability appears near the GBs firstly, and then expands to the bulk regions, shown by the black circle-marked regions in Figure 13(b2)-(b3). That is, the influence of GB (crystal defect) on the interface instability is more significant than the solidification parameters G and R, consistent with literature [52].

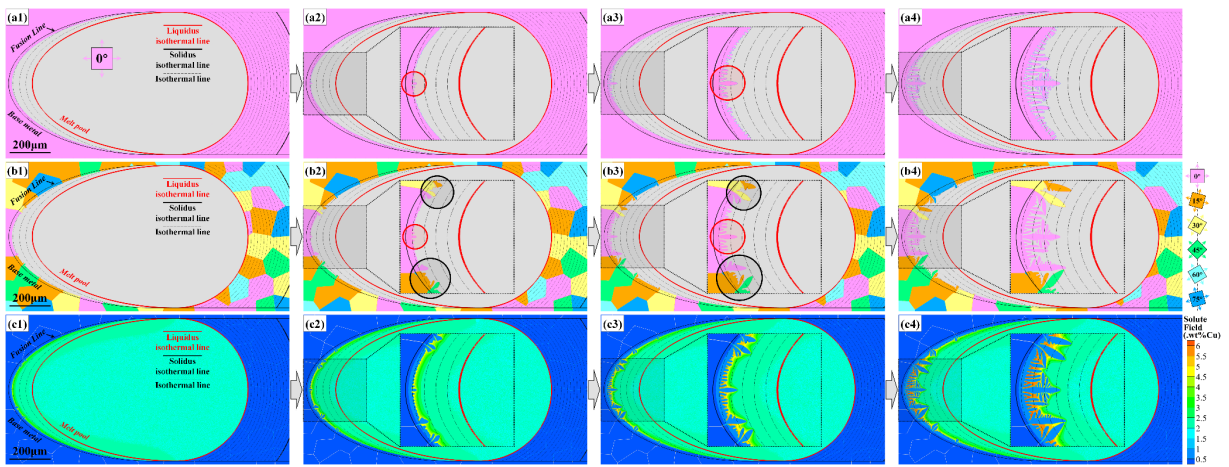

Figure 13. The solidification at the epitaxial growth stage: (a) the morphological evolution of single-crystal; (b) the morphological evolution of poly-crystal; (c) the solute-field evolution of poly-crystal. ($t_1 = 0.115s$; $t_2 = 0.143s$; $t_3 = 0.172s$; $t_4 = 0.201s$.)

**3.3.2. The competitive growth**

After the planar instability, the cellular prefer to grow along their <100> direction, to reduce the total free energy. The cellular with different PCOs will encounter and collide with each other, the process turns into the competitive growth stage, shown in Figure 14. The effects of variational solidification parameters and GBs on the microstructures at this stage will be discussed in the following.

It needs to be noted, in Figure 14(a2)-(a3), the gray dash lines reflect the growth directions of dendrite trunk, which are not parallel with the colorful arrows representing the PCOs. This phenomenon results from the fact that the growth direction of dendrite trunk is determined by the PCO and TGD simultaneously. When the PCO differs from the TGD, the growth direction will be between the PCO and TGD [53,54]. Moreover, near the GBs, the growth direction of dendrite trunk deviates from the PCO more obviously, indicating the significant effect of GBs on the microstructures.

As for competitive growth, on the one hand, the grains can be divided into two types, the Favorable Oriented (FO) grains and Unfavorably Oriented (UO) grains, based on the theory proposed by Walton and Chalmers (WC theory) [55]. Specifically, the grains having small intersection angle between the PCO and TGD are called the FO grains, while the grains having larger intersection angle than the FO grains are called the UO grains. As solidification goes on, the FO grains keep growing by blocking the UO grains. Finally, the FO grains grow to the pool center, while the FO grains are eliminated, shown in Figure 14(a1)-(a4). As a result of competitive growth, the directional evolutions of GBs are dominated by the TGD, consistent with literature [56]. As shown by the dotted lines in Figure 14(a4), the directions of GBs are from the fusion line to pool center.

On the other hand, besides the definitions of the FO and UO grains, the grains can also be distinguished based on the dendrite trunk, the converging GBs and diverging GBs [38]. When the <100> dendrite trunks are converging, the dendrite tips of the UO grains hit the FO grains slightly behind their tips and become blocked, resulting in the converging GBs coincide with the orientation of the FO grains. By contrast, when the <100> dendrite trunks are diverging, a liquid region continuously opens between the two dendrites, extending the secondary arms into the gap. Then the secondary arms initiate the tertiary arms parallel to the primary trunks, and some of the tertiary become new primary trunks. The onset of the tertiary dominates the evolution of the diverging GB, and the direction of diverging GB is approximately a bisector of the <100> directions of the two dendrites [38]. In the melt pool, the variational conditions make the TGDs vary with space and time. Hence the FO grains and UO grains might switch to each other during solidification. Since the converging GBs coincide with the orientation of the FO grains, the directions of converging GBs are circuitous, shown by the black dotted lines in Figure 14(a4). By contrast, the directions of diverging GBs, determined by the respective <100> directions of the two dendrites, remain relatively straight, shown by the red dotted lines in Figure 14(a4).

Furthermore, the space-time variational parameters make the competition in melt pool more complex, bringing the special microstructures. In Figure 14(a4), the pink grains grow from the tail and survive to the end, along the weld direction. The continuous grains, along the weld direction and around the centerline, are called axial grain structures, which appear under relatively low speed and heat input [57]. In current paper, the weld speed is 1.0mm/s and the power is 1200W, belonging to low speed and heat input. Hence the axial grain structures are observed in the pool. By contrast, the axial grain structures do not appear under higher weld speed, in literatures [15] and [16]. The formation of the axial grain structures can be regarded as one kind of competitive growth, between the grains along the axial direction and radial direction. The detailed investigation will be carried out in the future.

In conclusion, the variational parameters across the pool affect the solidification evolution significantly, including the directional evolutions of the converging GBs and diverging GBs, as well as the formation of axial grain structures near the centerline.

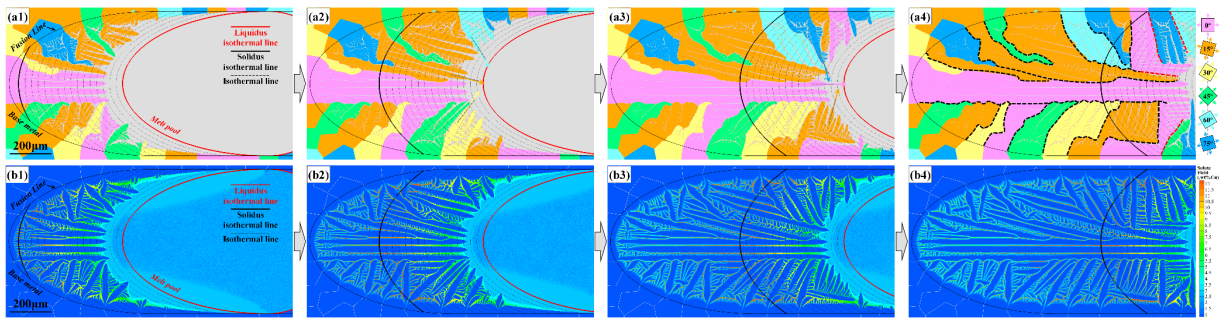

Figure 14. The poly-crystal solidification at the competitive growth stage: (a) The evolution of grain; (d) The evolutions of sub-grain and solute-field. ($t_1 = 0.545s$, $t_2 = 0.831s$, $t_3 = 1.118s$, $t_4 = 1.405s$.)

## 4. Summary and outlook

In this paper, the microstructure evolutions under space-time variational solidification conditions in a melt pool are carried out through the multi-scale simulation. Firstly, the GR model, using the variational G and R, and the TF model, using the variational temperature field T(**r**, t), are used to connect the PF equations with the macroscopic thermal parameters. The results indicate the TF model could represent the variational parameters across the pool better than the GR model. Then, the single-crystal solidification and poly-crystal solidification at the whole region of pool are carried out via the TF model. The influences of solidification parameters and crystallographic parameters on the microstructures is investigated. The following conclusions can be drawn from the study:

(1) For single-crystal solidification at the local region, the non-periodic evolutions indicate the necessity of considering the variational parameters, while both the GR model and TF model can reflect the variabilities. On the other hand, the TF model can represent the variational parameters better than the GR model.

(2) For solidification at the whole region, the evolutions of parameters G, R, G*R and G/R demonstrate the space-time variabilities of solidification parameters across the pool. Moreover, the parameters at the tail of pool remain stable while the parameters at the bottom of pool are unstable. As time goes on, the parameters at different locations tend to be similar.

(3) At the epitaxial growth stage, the planar instability is induced by interface energy and its anisotropy, where the interface energy is determined by the excess free energy of the S/L interface. In the single-crystal solidification, the excess free energy is induced by solute segregation, which is determined by solidification parameters. Hence the effect of solidification parameters is greater than crystallographic parameters. In the poly-crystal solidification, due to their higher free energy, the influence of GBs is greater than solidification parameters. That is, for the planar instability, the effect of GB (crystal defect) is greater than solidification parameters, while the effect of solidification parameters is greater than crystallographic parameters.

(4) At the competitive growth stage, on the one hand, most grains follow the WC theory. The FO grains keep growing by blocking the UO grains. Eventually, the FO grains grow to the center while the FO grains are eliminated. On the other hand, the GBs can be classified as the converging GBs and diverging GBs. The converging GBs coincide with the orientation of the FO grains, while the diverging GBs evolve along the bisector of the <100> directions of the two dendrites. The variational parameters cross the pool make the FO grains and UO grains switch to each other, resulting in the converging GBs are circuitous while the diverging GBs are relatively straight.

(5) The axial grain structures are observed in this study, under the relatively low speed and heat input. The formation of axial grain structures can be regarded as the competition between the grains along the axial direction and radial direction. This special phenomenon also indicates the necessity of considering the space-time variational parameters in the melt pool.

The investigations in this paper demonstrate the space-time variabilities of solidification conditions in the melt pool. Moreover, the variational conditions across the pool influence the microstructure evolution significantly. Hence the simulation of single-crystal solidification at a local region only makes sense at the beginning. As times goes on, the space-time variational conditions should be considered. Meanwhile, the quantitative PF model can simulate the microstructure evolution under the variational conditions accurately, which has a great potential for investigating solidification dynamics in the melt pool.

It should be noted, the investigations in this paper have some limitations:

(1) At the planar crystal growth stage, due to the shape of melt pool, the interface is not real planar. Hence the theories of the planar growth should be modified for melt pool solidification.

(2) In section 3.2.2, we attribute the excess free energy of the S/L interface to the solute segregation. Two points need to be explained:

- In current PF model, Gibbs-Thomson coefficient $\Gamma = \gamma_{sl}T_f/(\rho_s L_f)$ is constant, corresponding to constant interface energy $\gamma_{sl}$. The $\gamma_{sl}$ here is the average from experimental measurements. In the PF simulation, the solute segregation makes the state of interface correspond to the different points in the phase diagram, representing different Gibbs energies of the interface. Hence, the discussion that the excess free energy is determined by the solute concentration is reasonable.
- In current PF model, the solution is regarded as the ideal dilute solution, which means the free energy does not include the bond energy between the components. Thus, the PF model needs to be modified. Fortunately, the model formulated from the grand-potential functional [58,59] could consider the bond energy between the components, will be used in the future.

**Acknowledgments**

This work is supported by the fellowship of China Postdoctoral Science Foundation (2021M692040), Fundamental Research Funds for the Central Universities (NP2016204), National Key Research and Development Program (2017YFB0305301) and National Natural Science Foundation of China-Excellent Young Scholars (51922068).

The authors acknowledge the technical support from the Center for High Performance Computing at the SJTU, as well as the financial support by the Priority Academic Program Development of Jiangsu Higher Education Institutions (PAPD). The authors acknowledge Dr. Damien Tourret in the IMDEA Materials Institute for the help in PF theory and solidification theory.

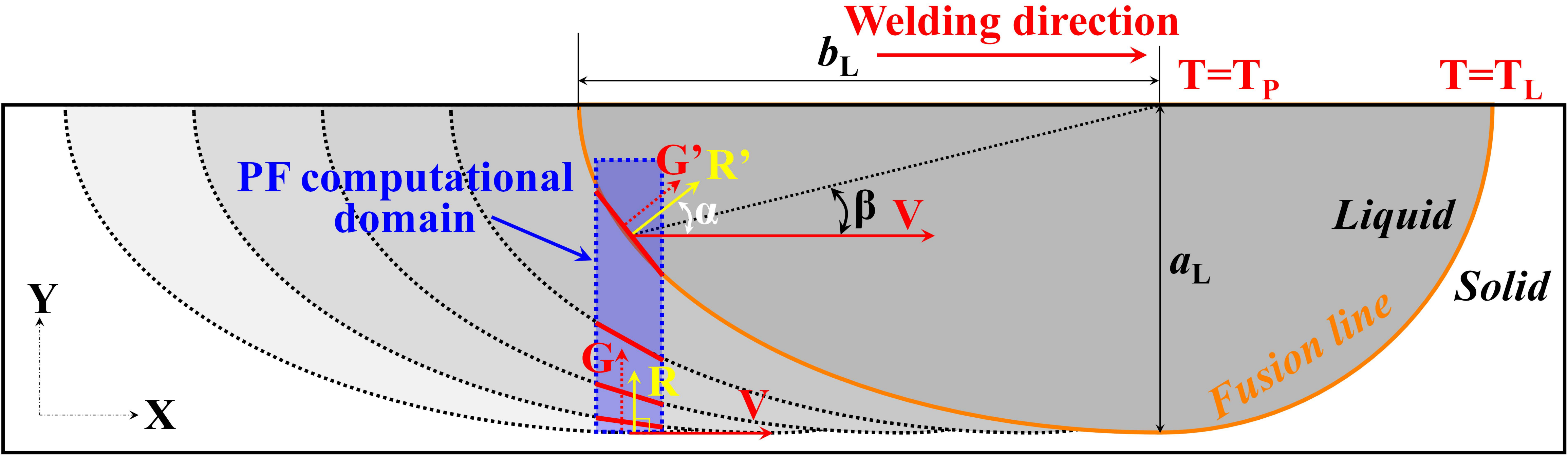

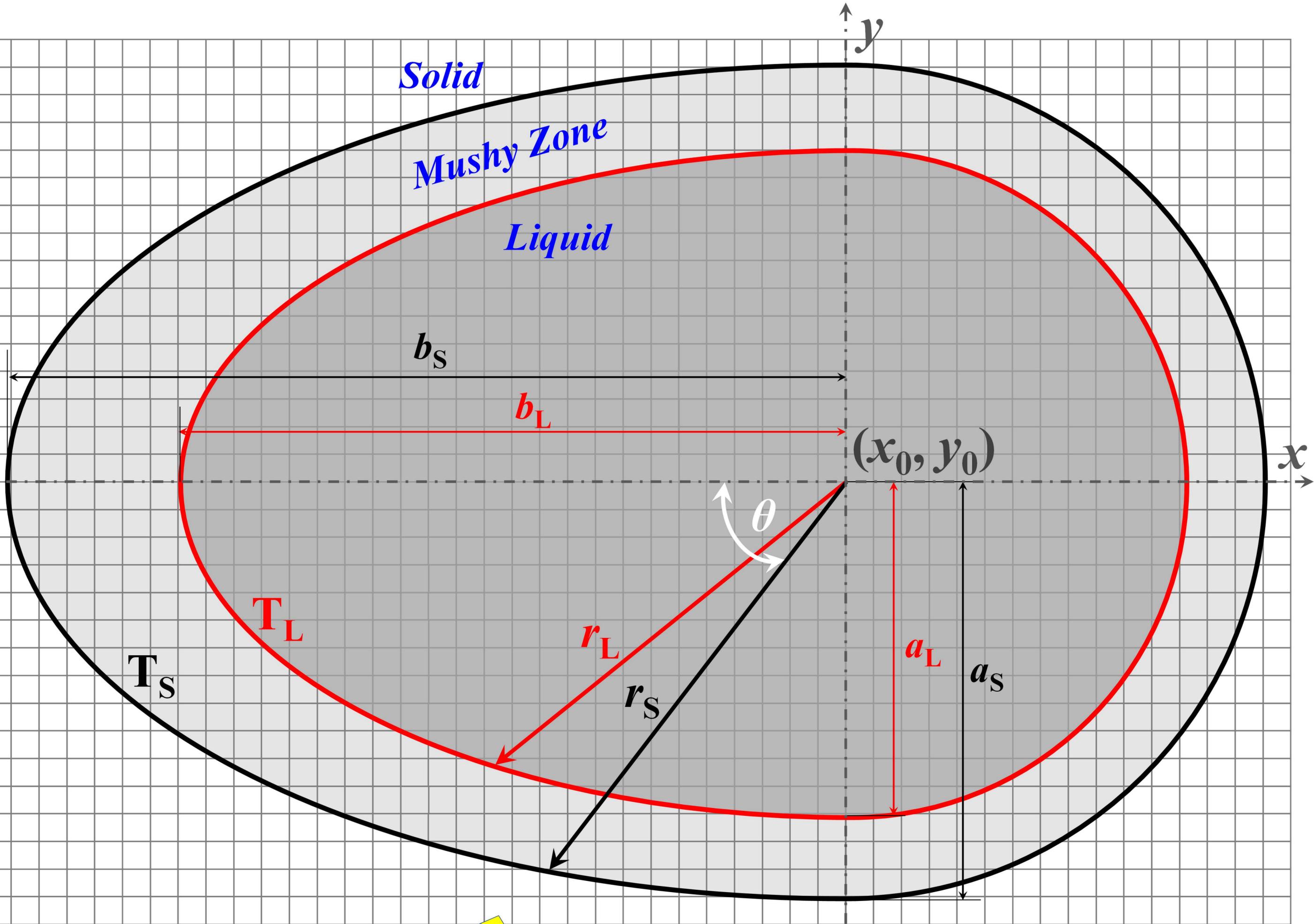

The motion of melt pool in PF simulation
Time steps of each space step: Δx / weld speed

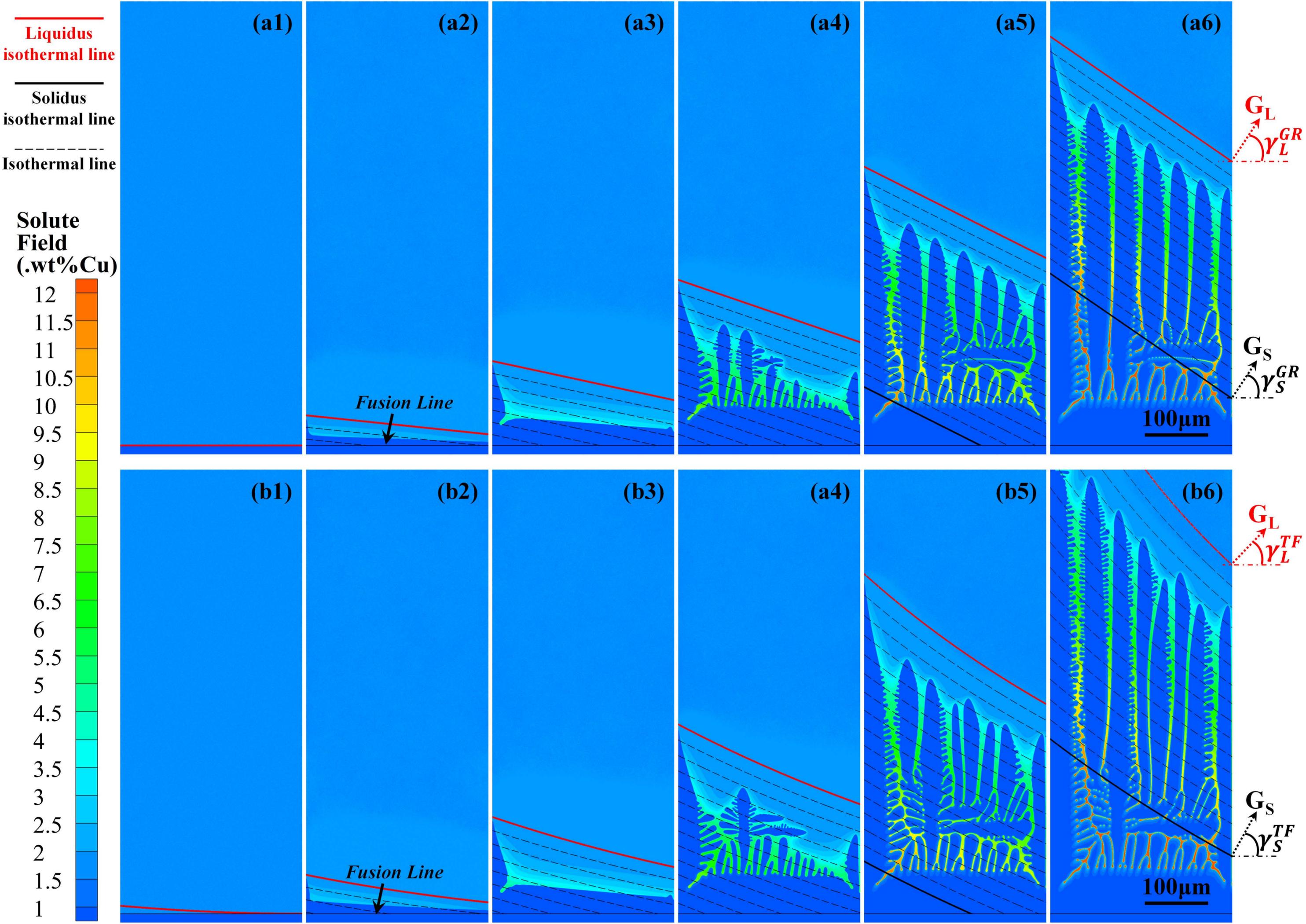

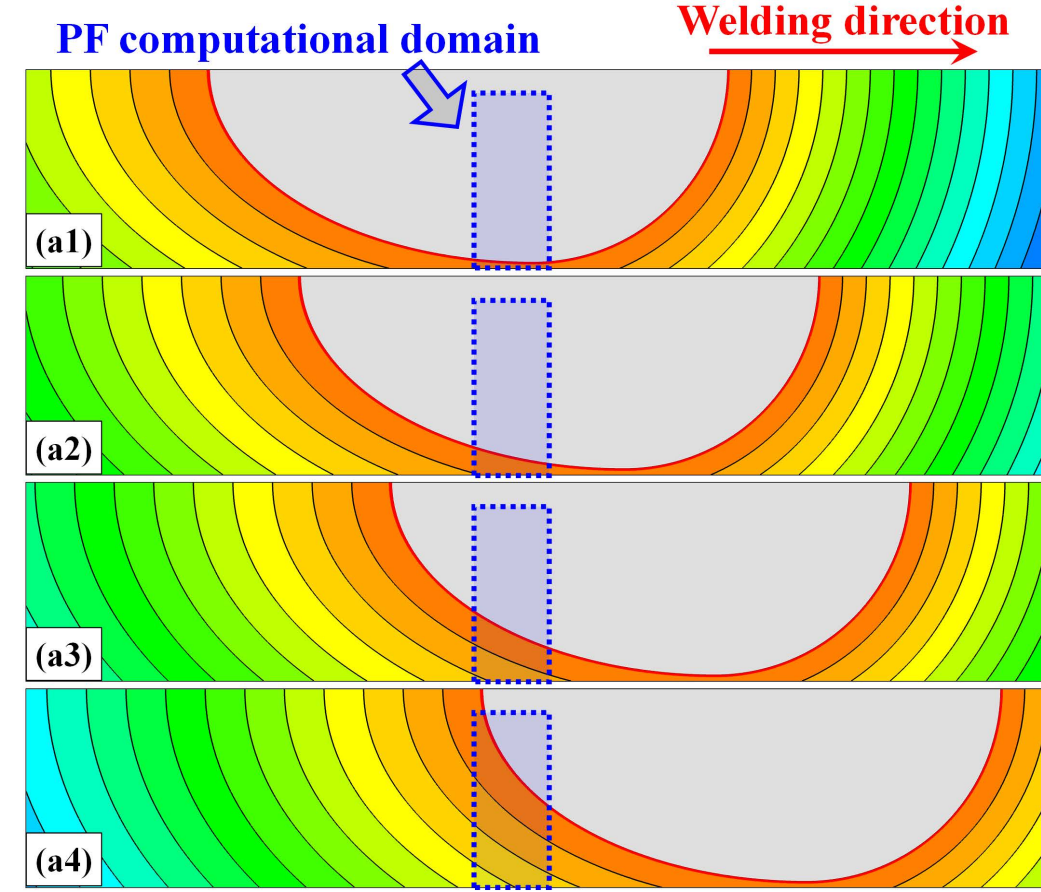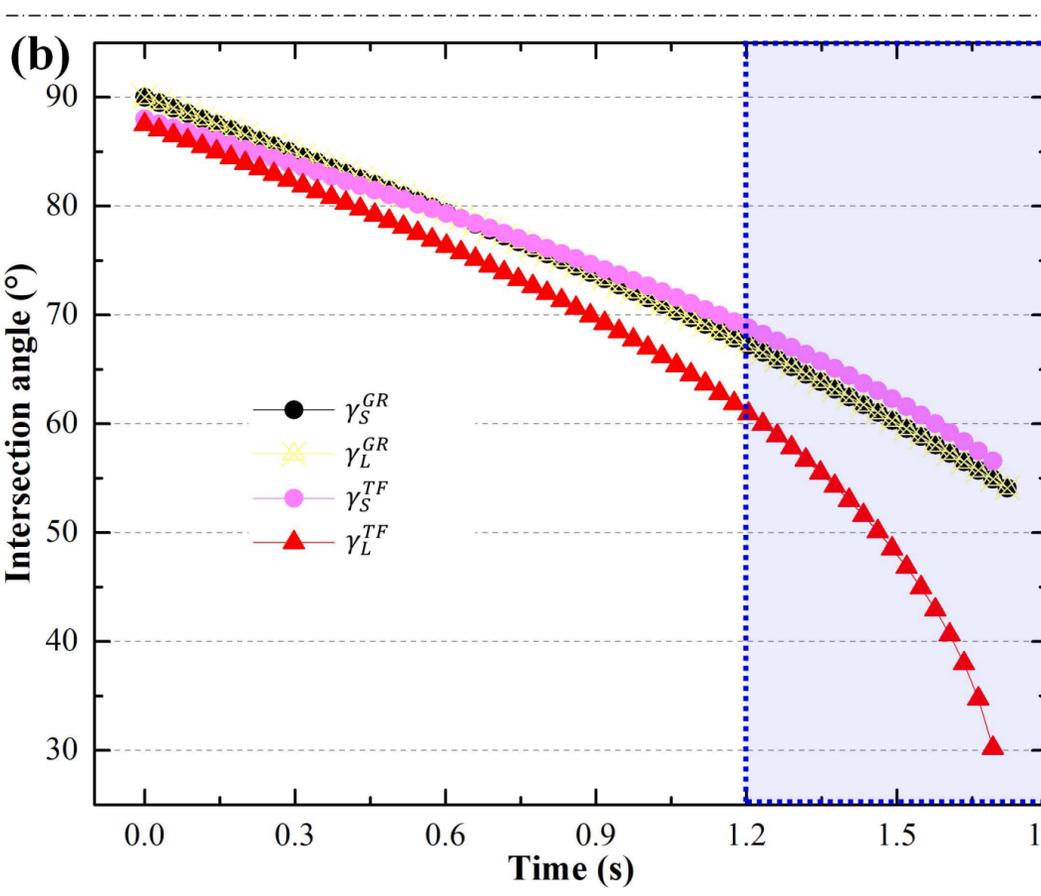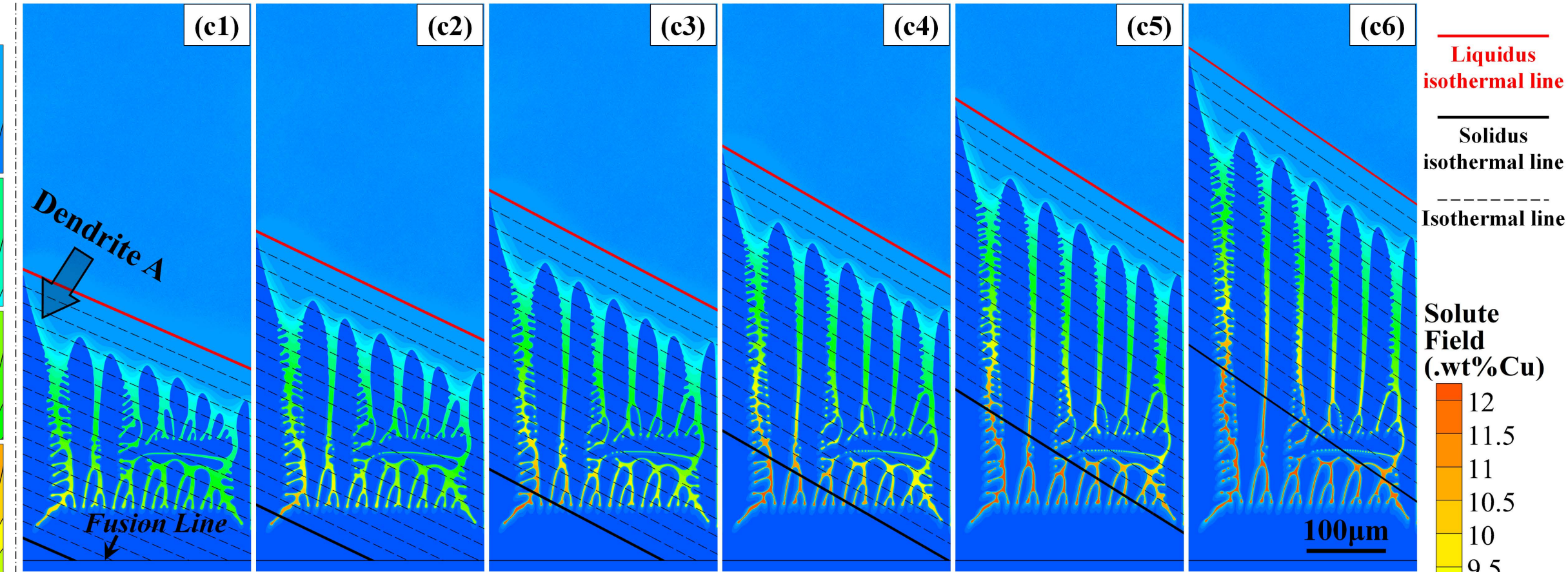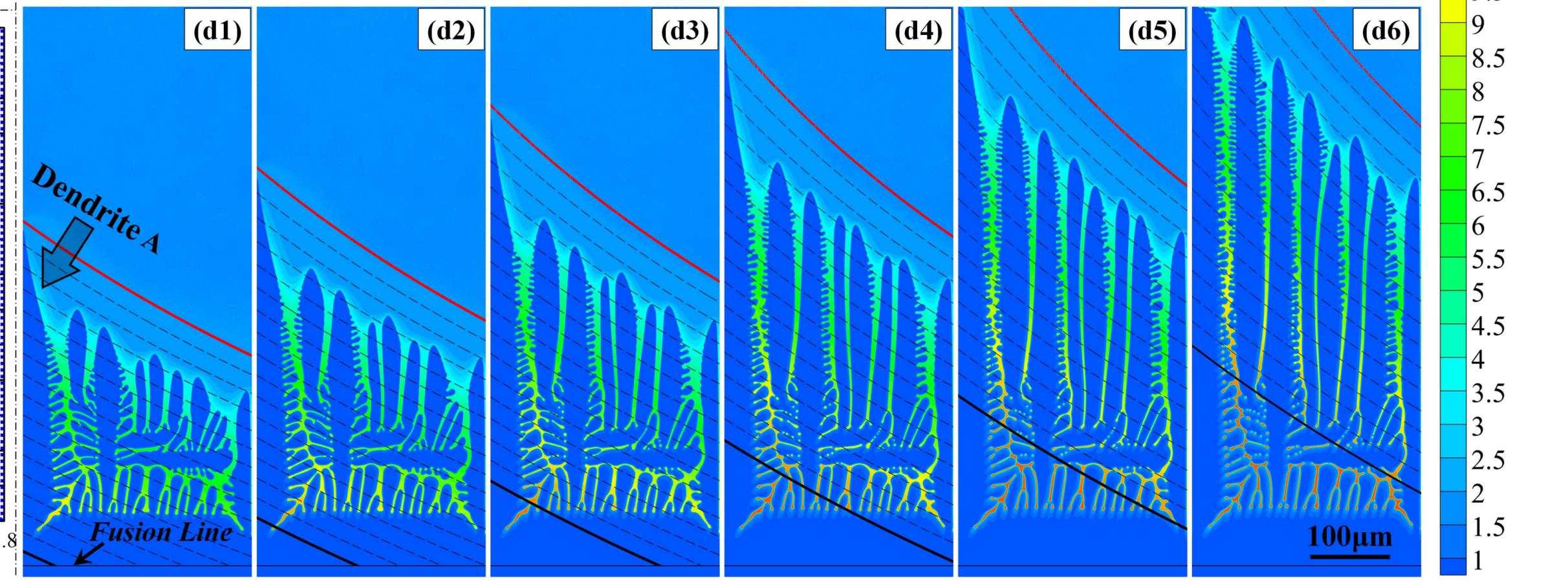

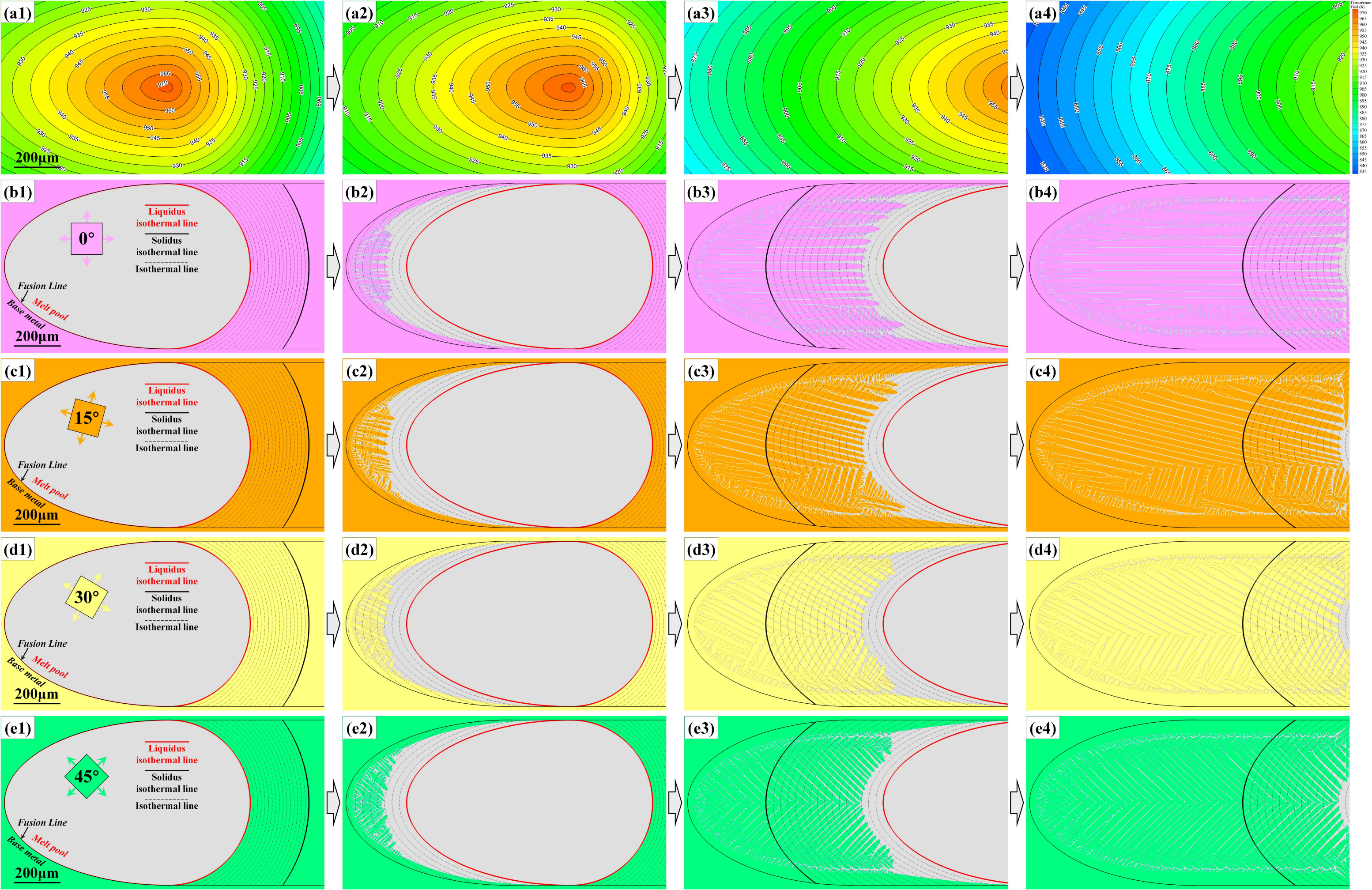

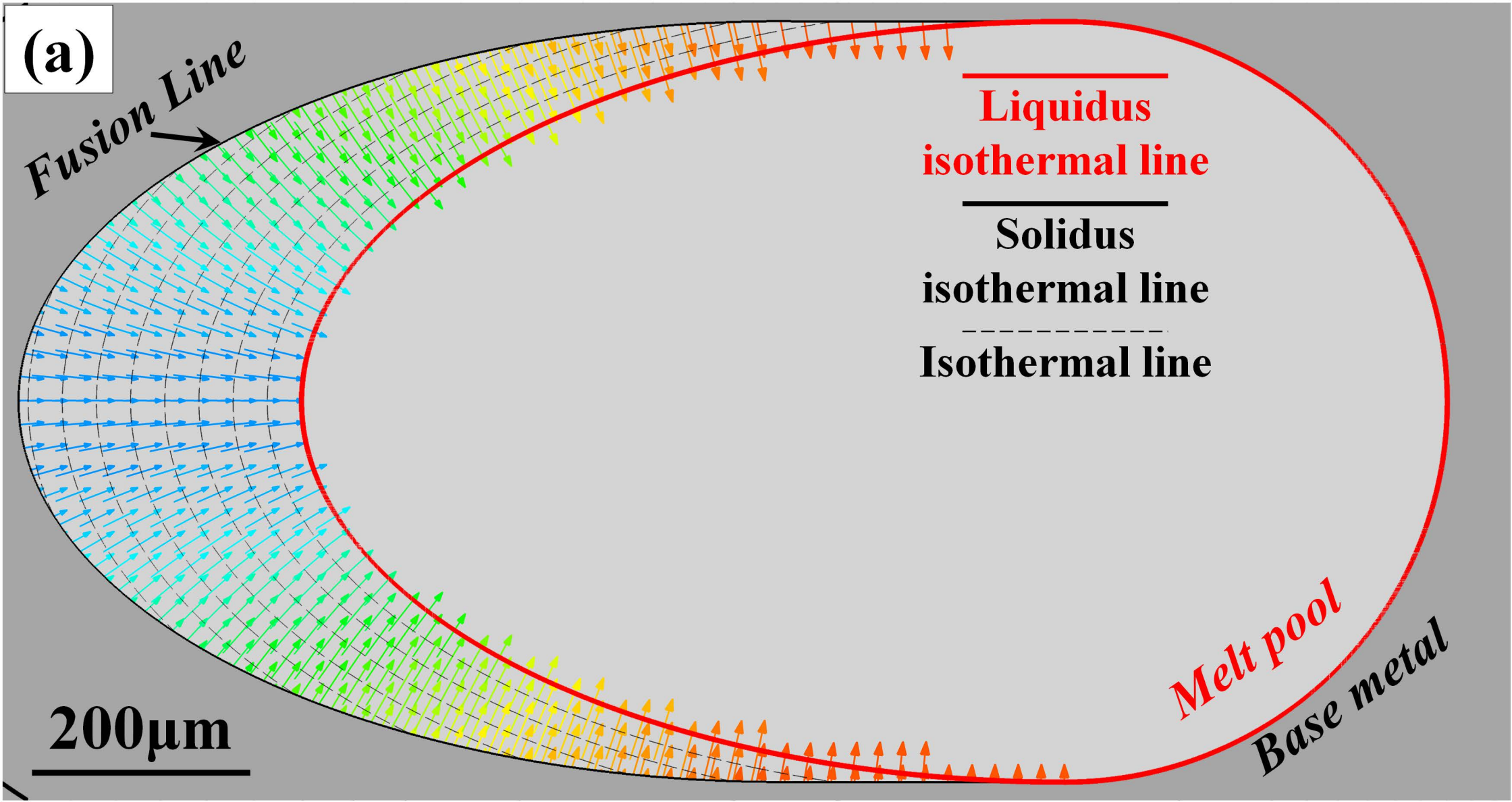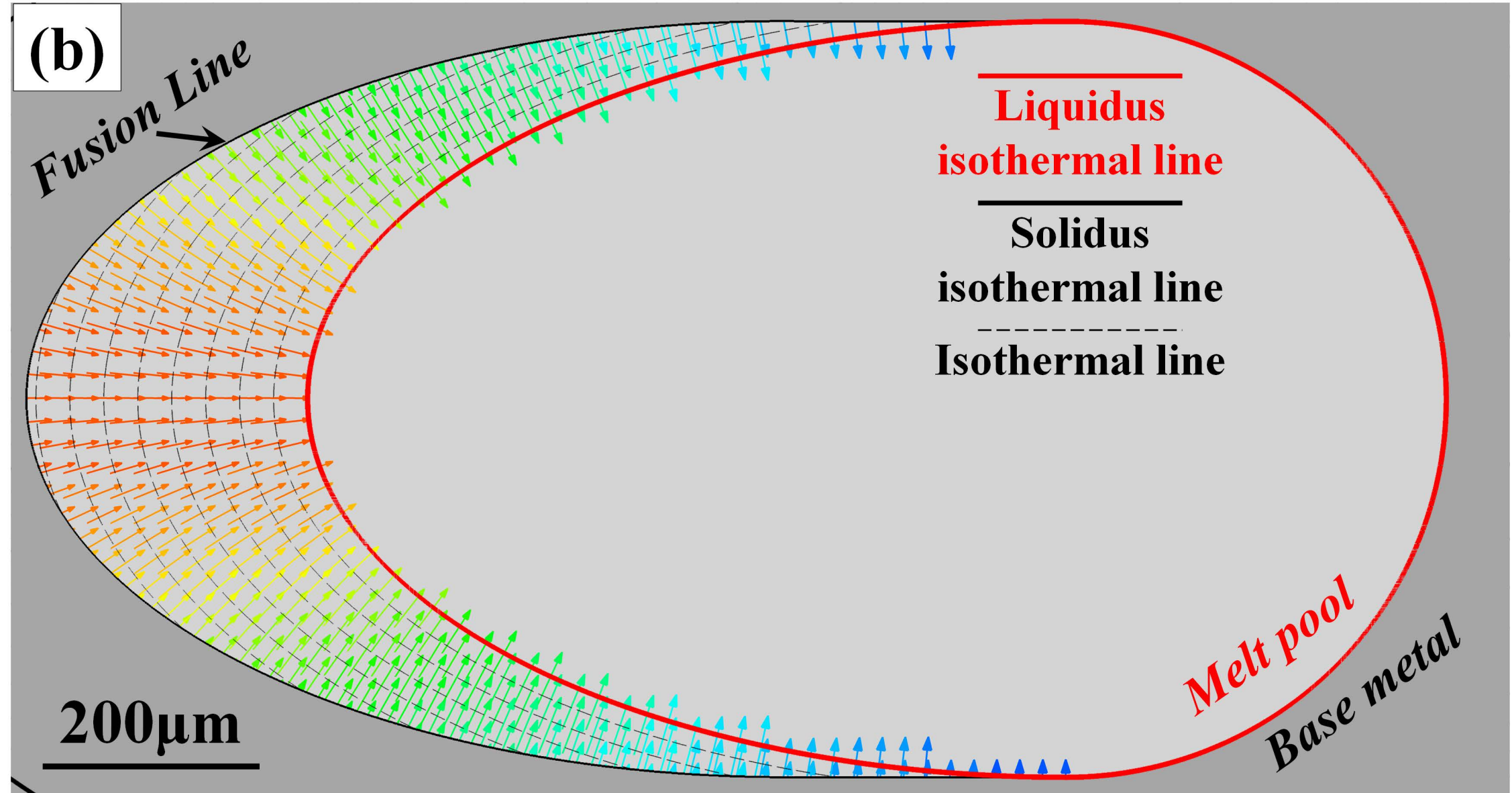

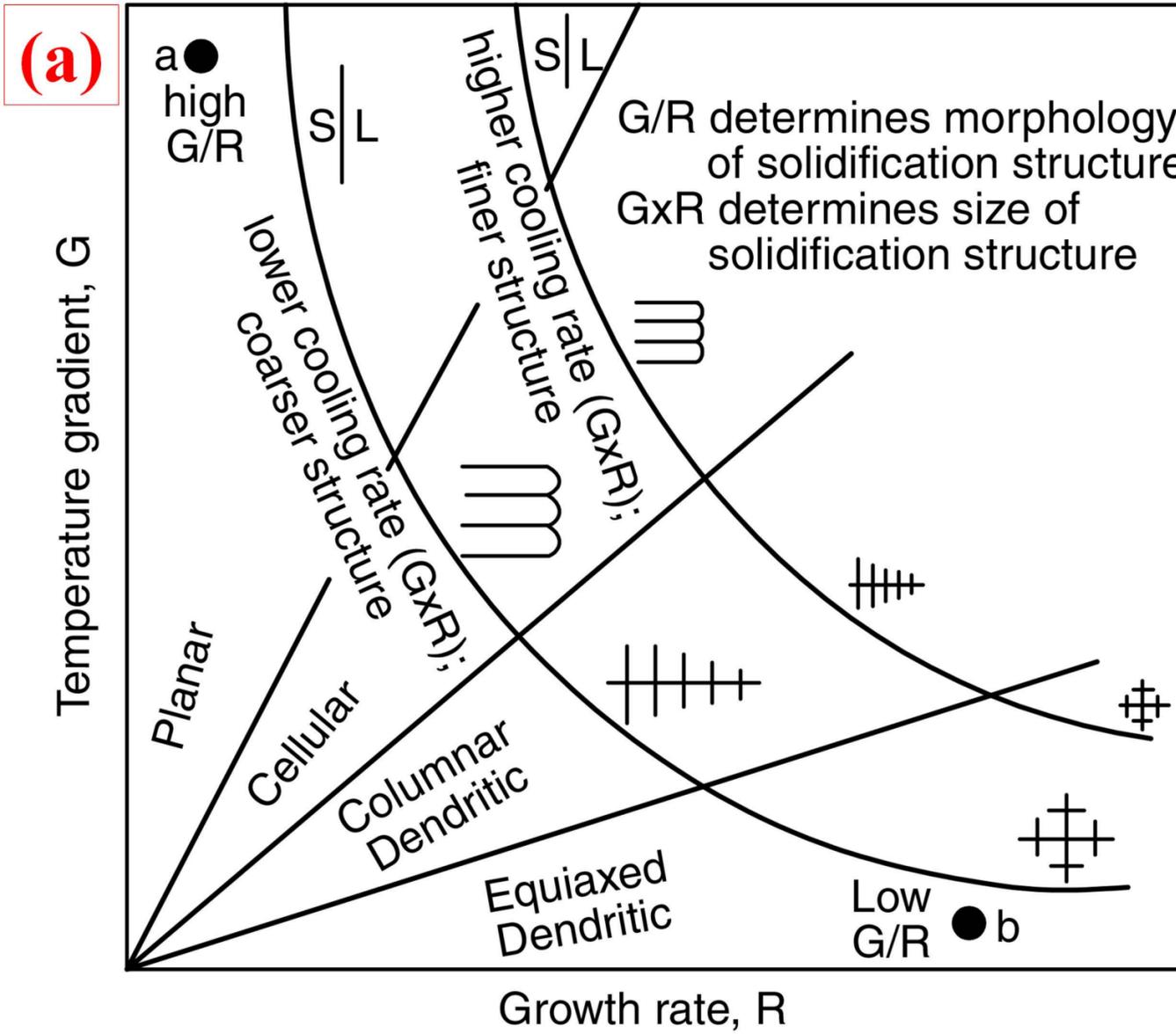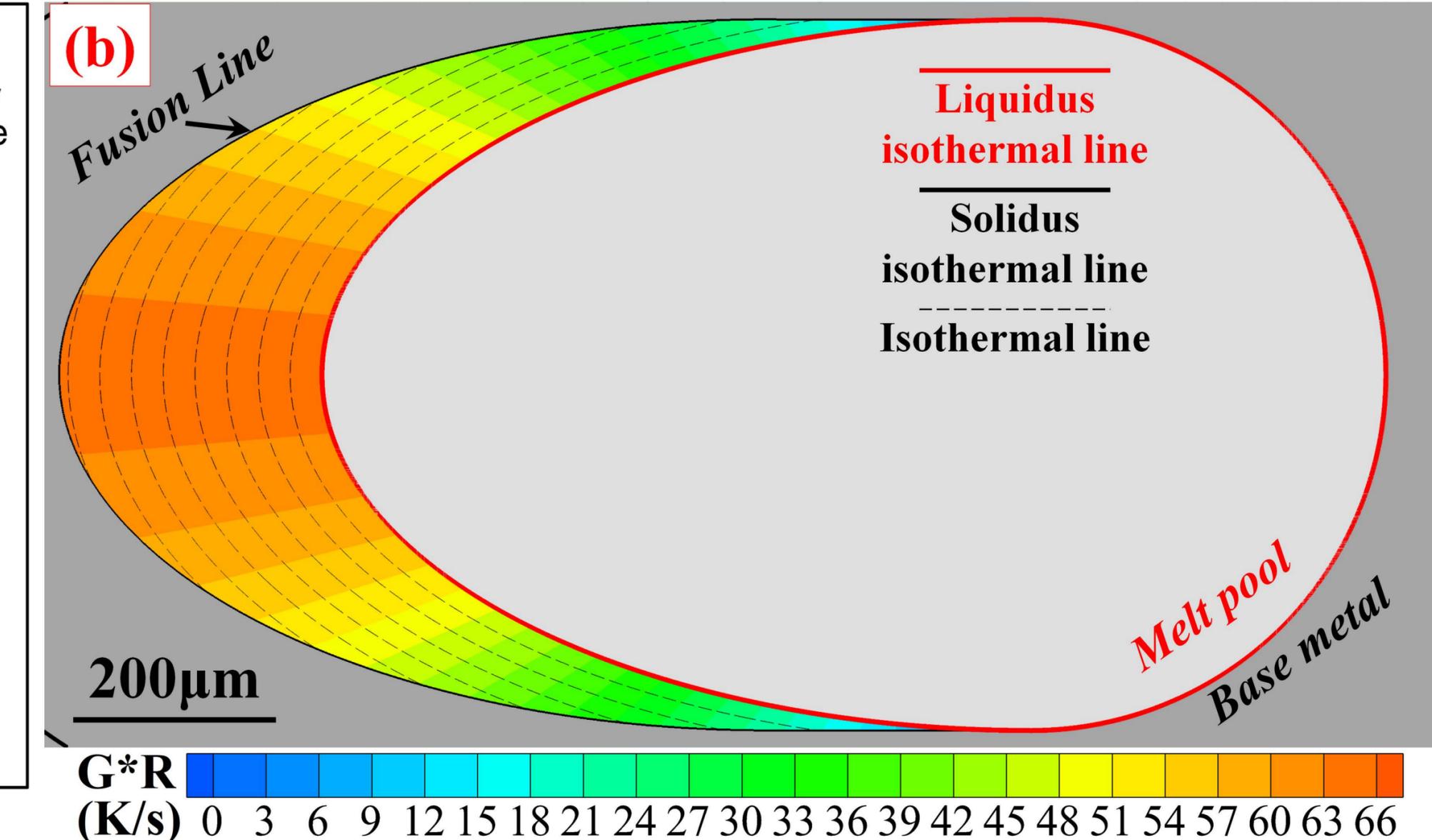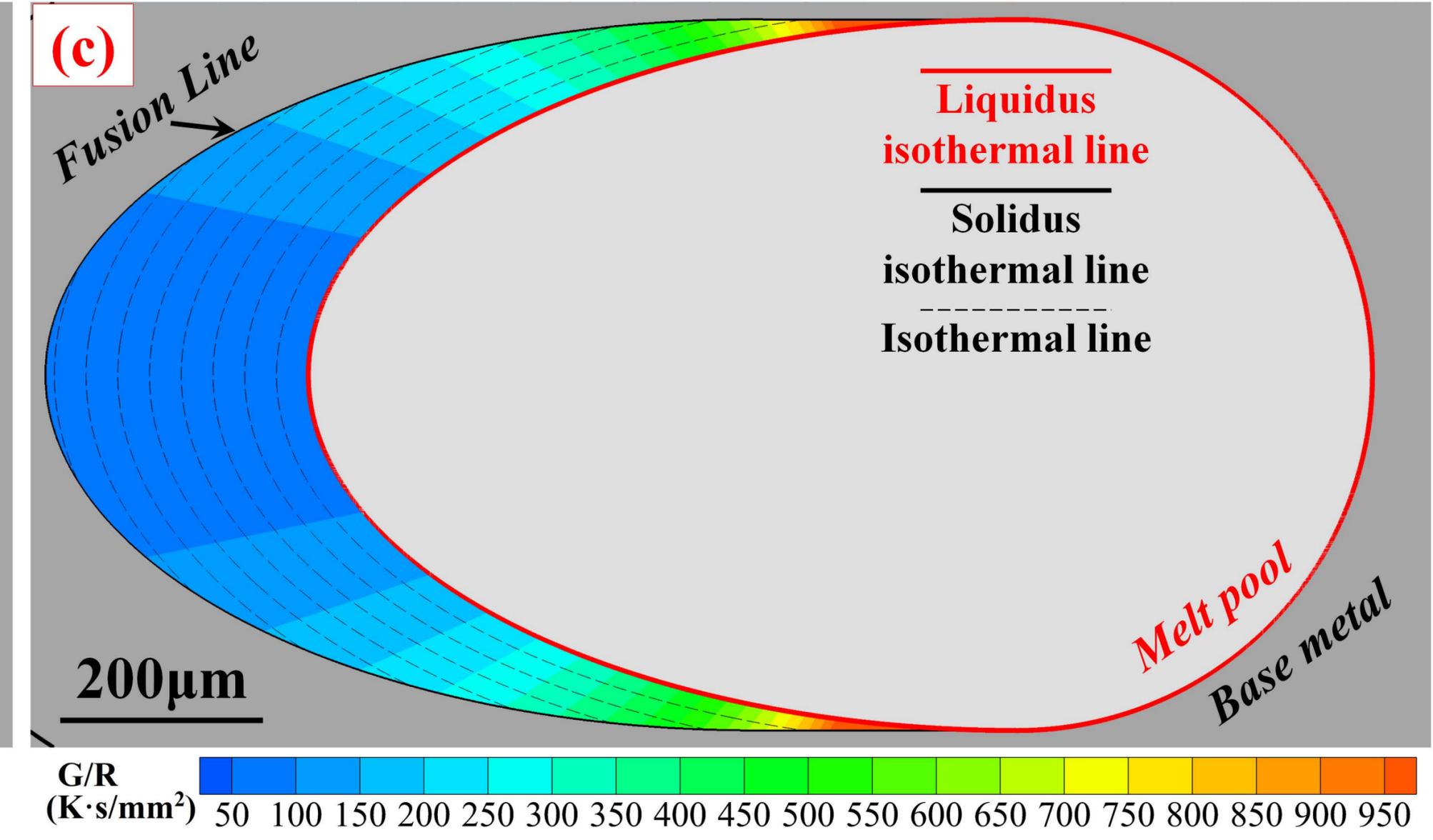

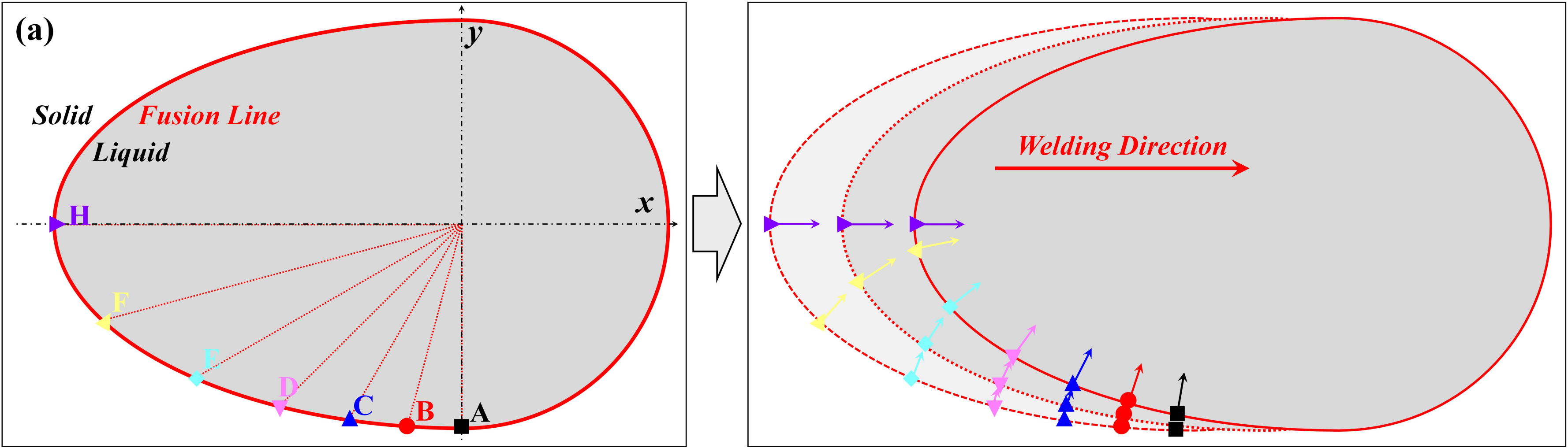

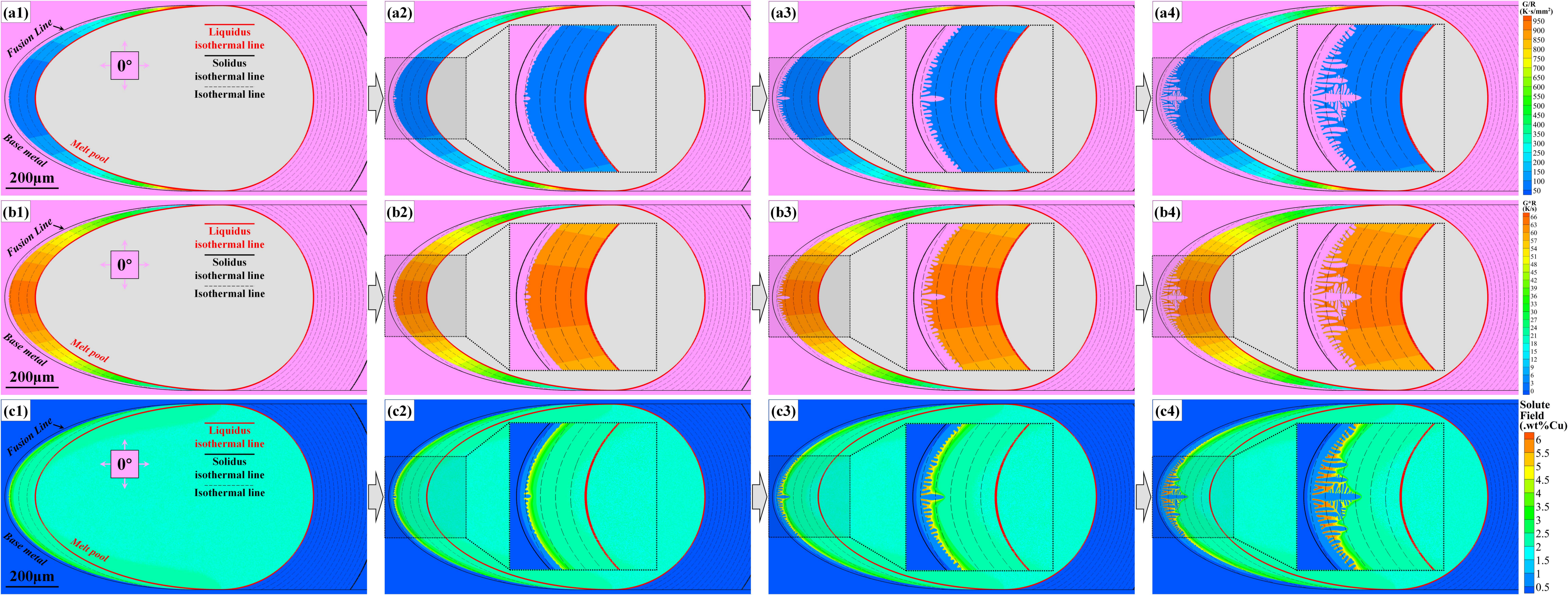

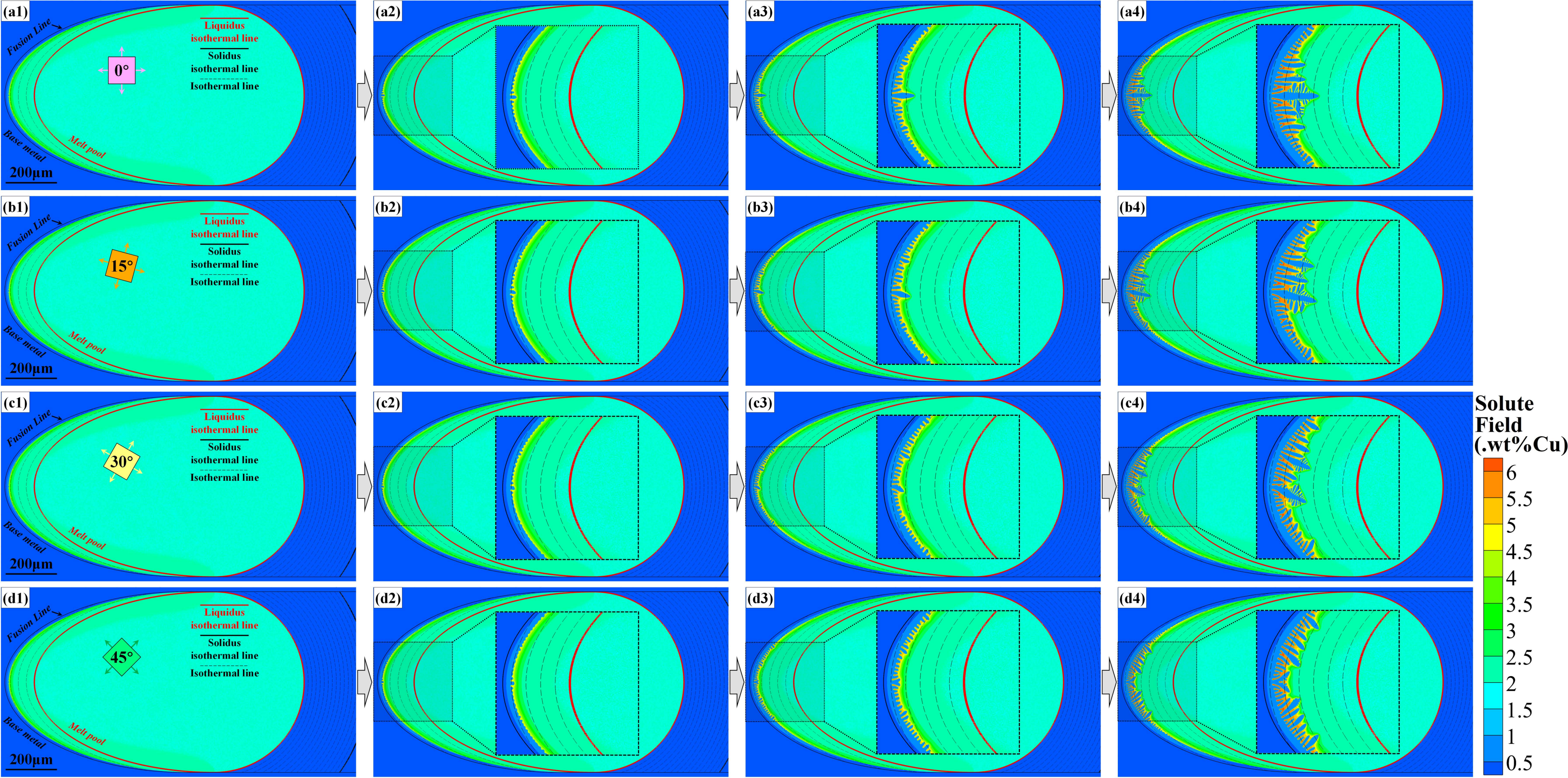

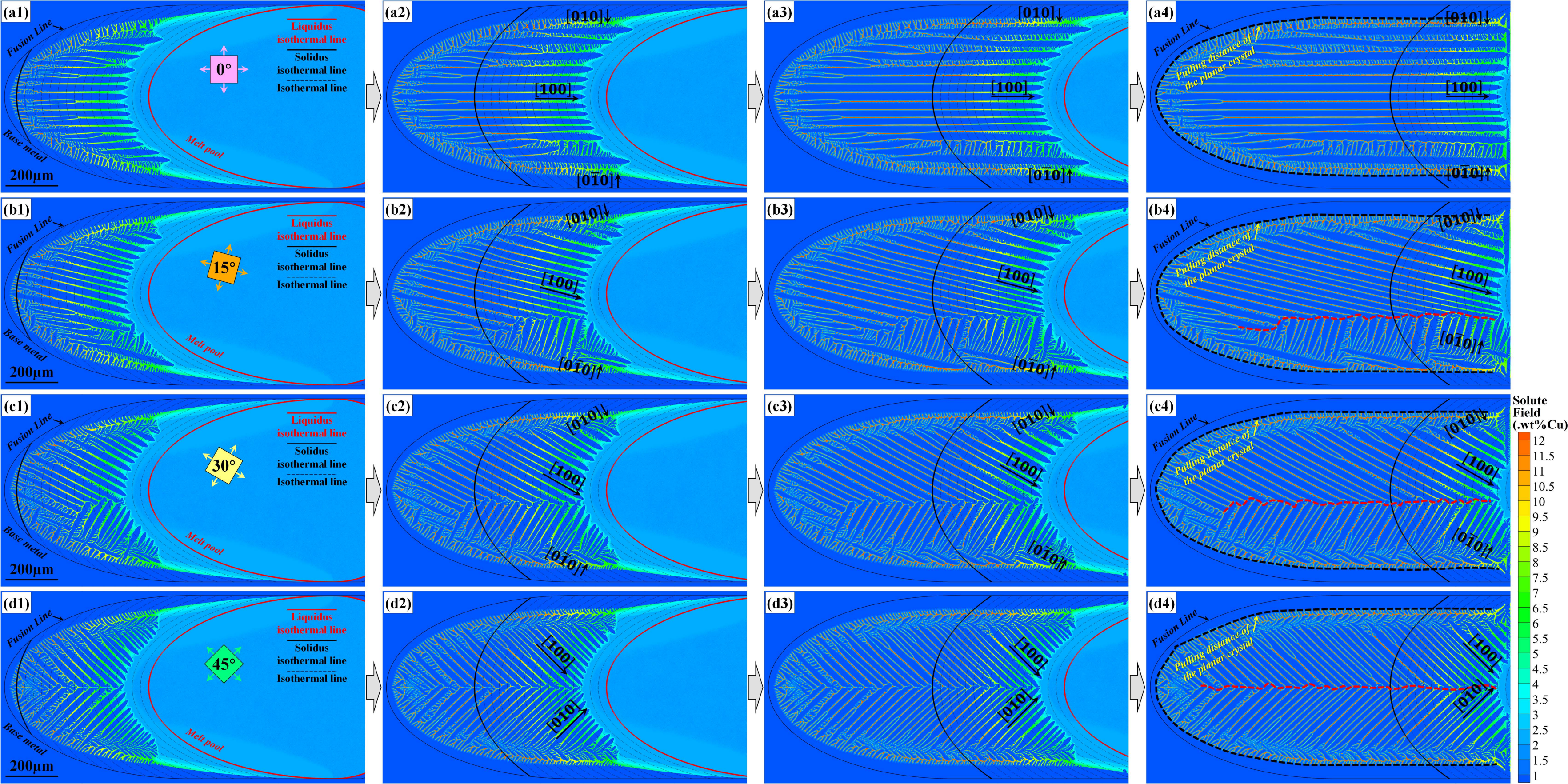

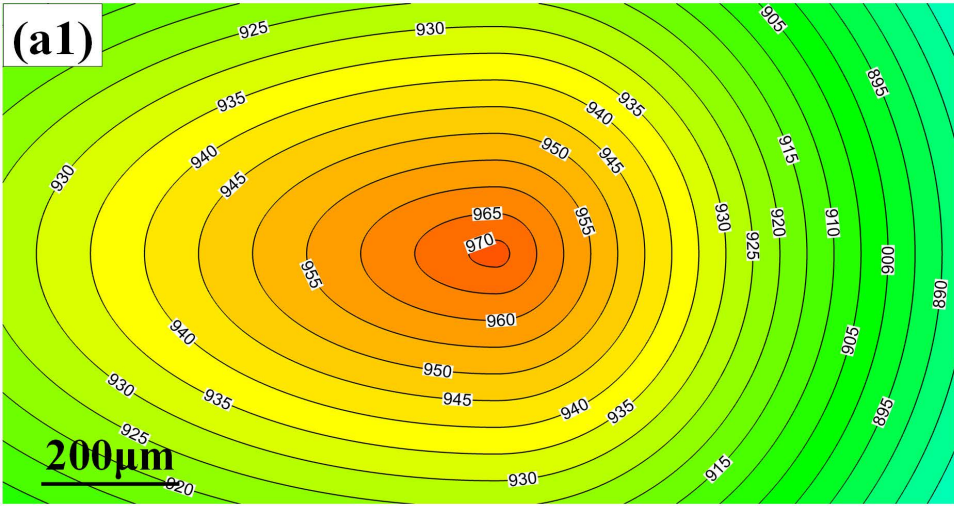 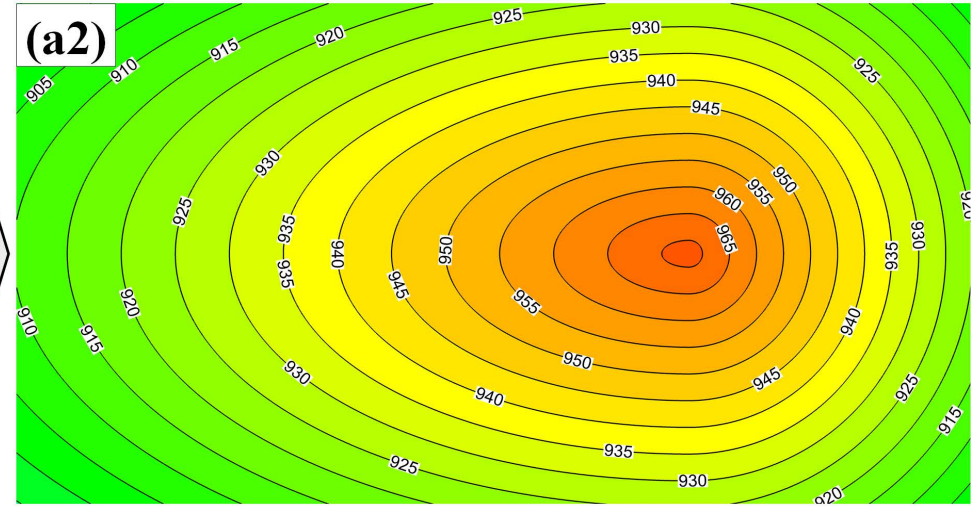 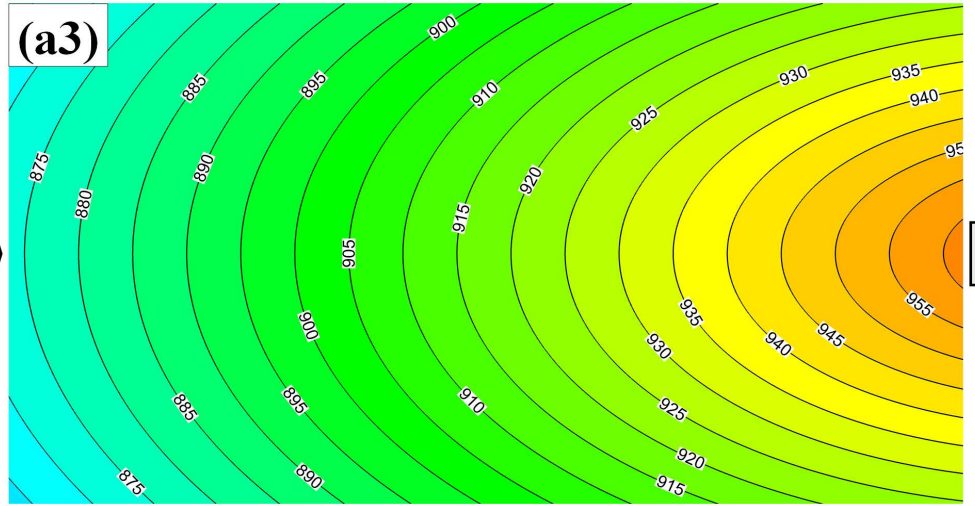 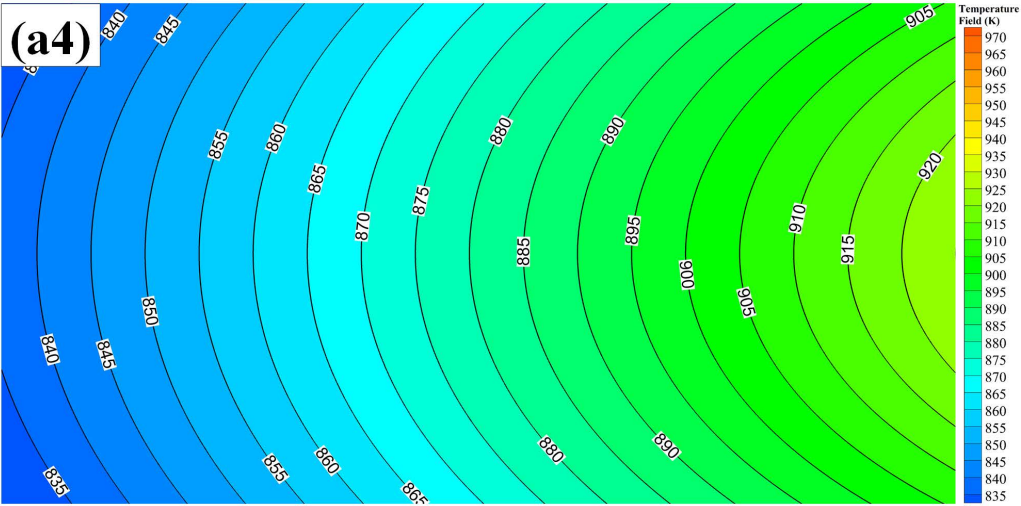
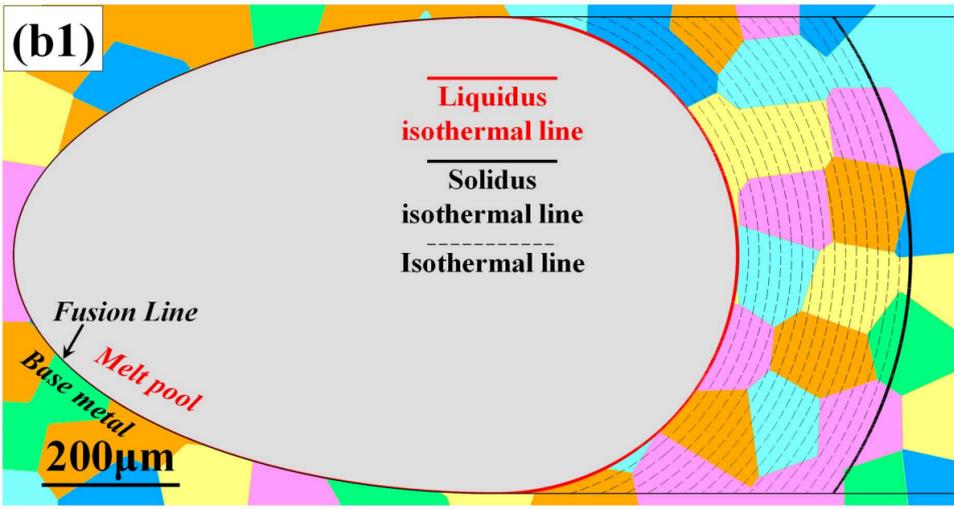 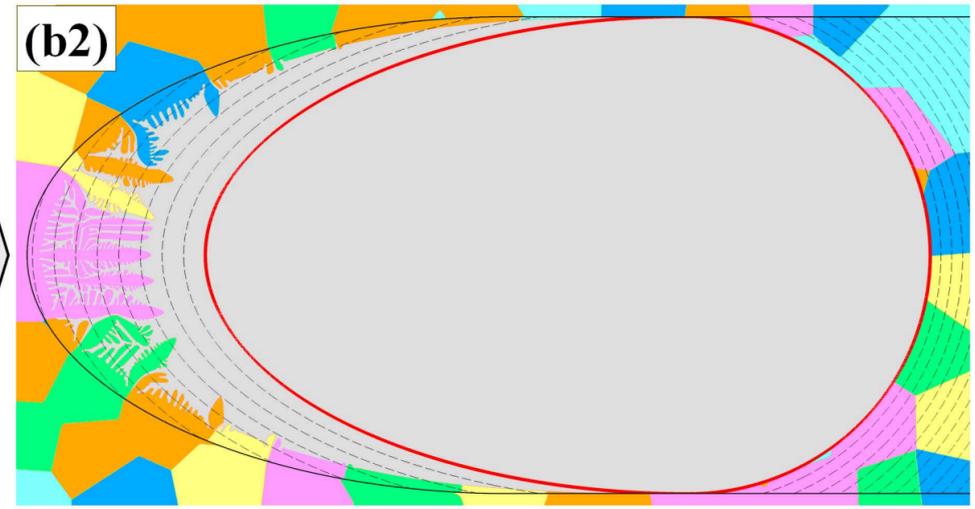 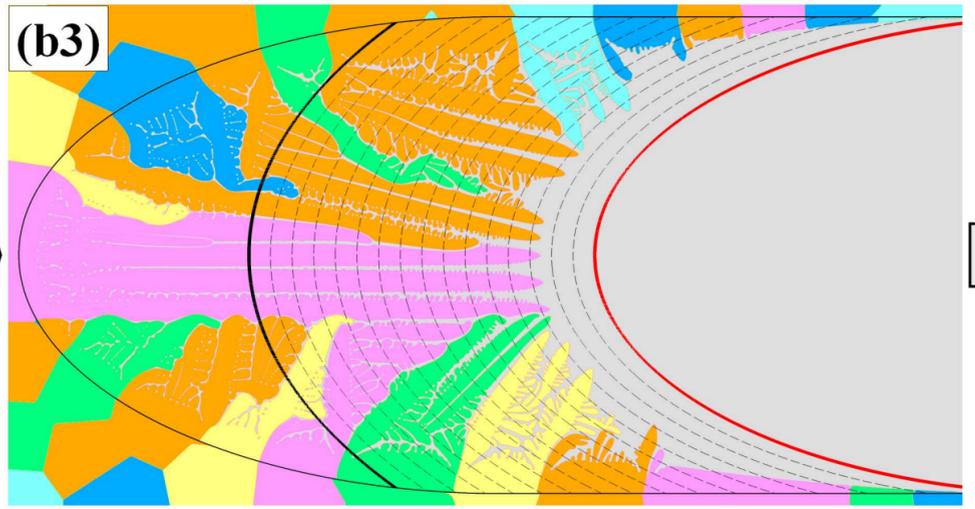 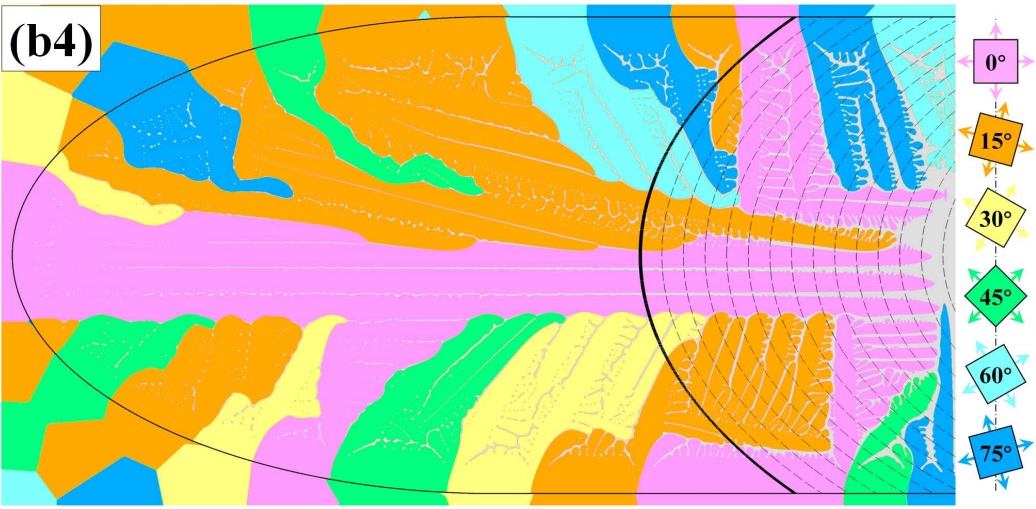
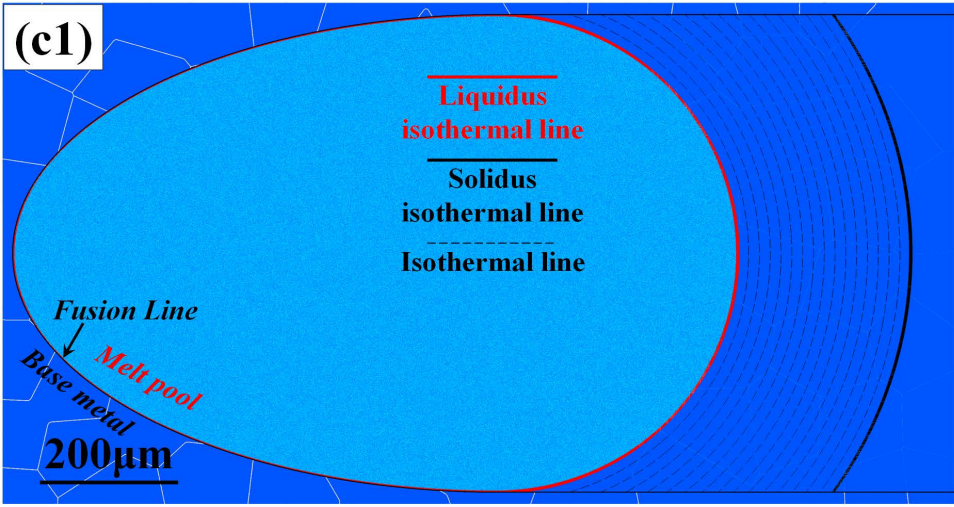 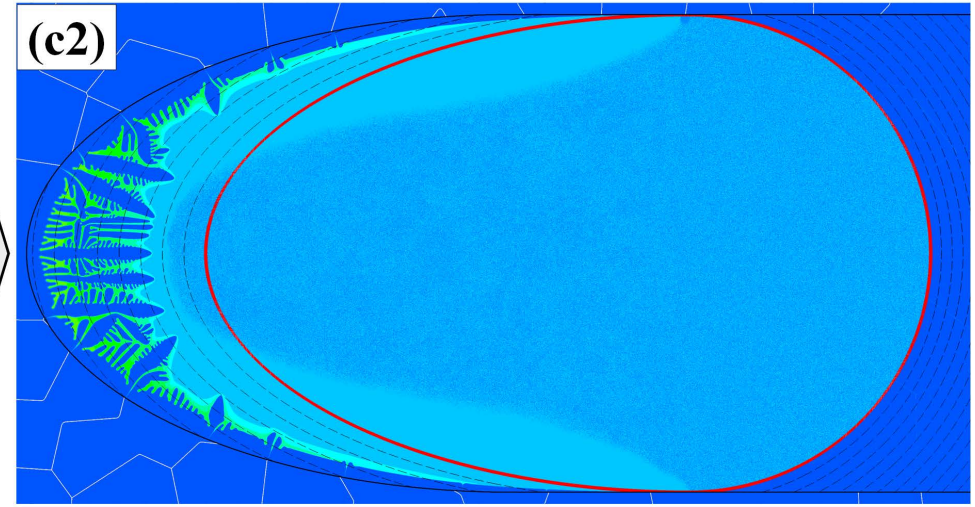 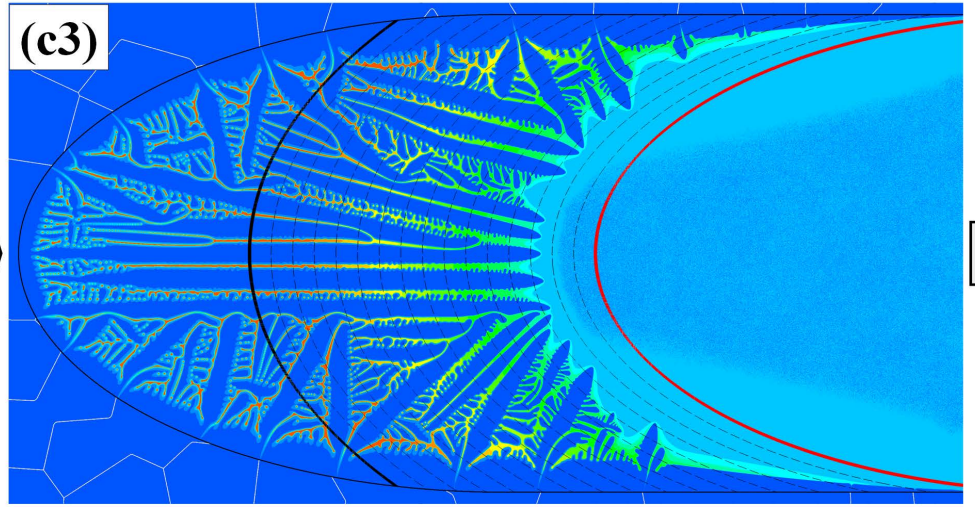 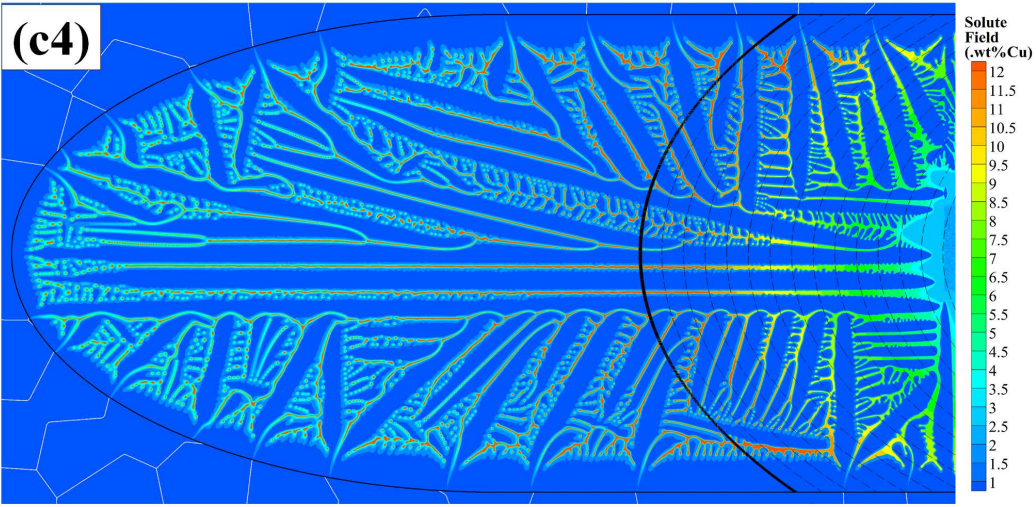

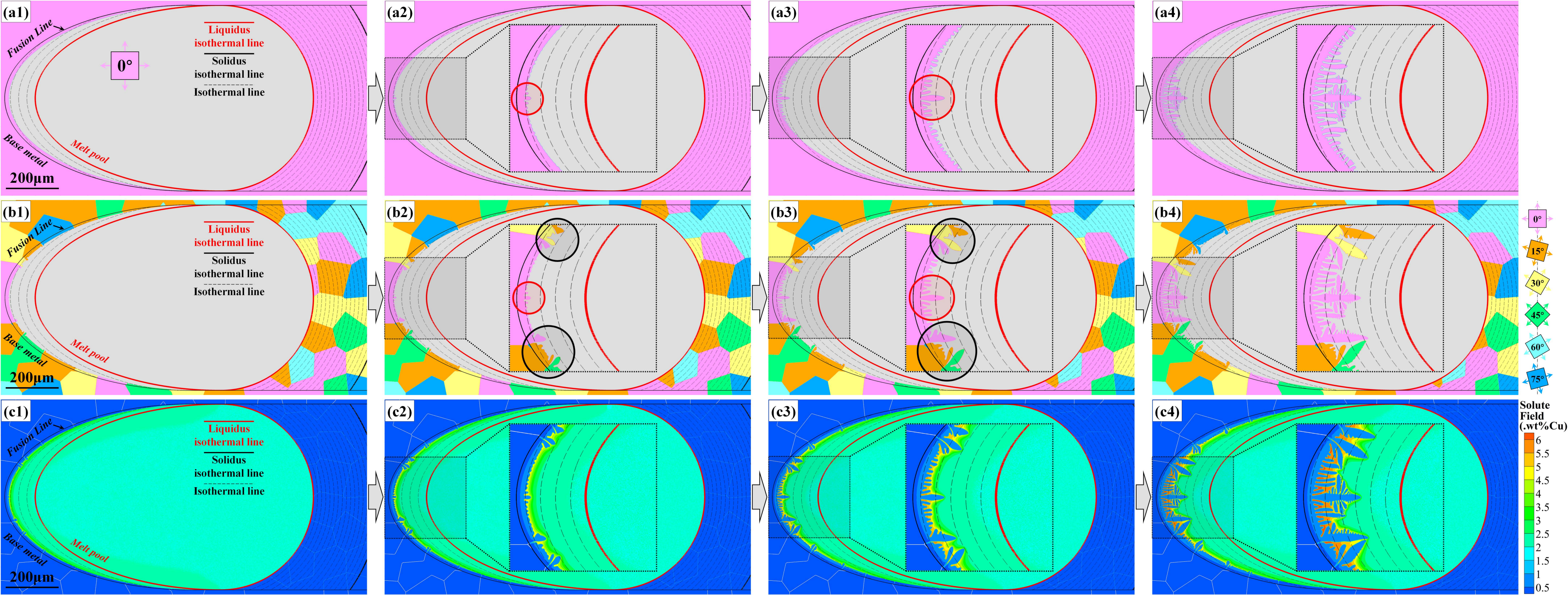

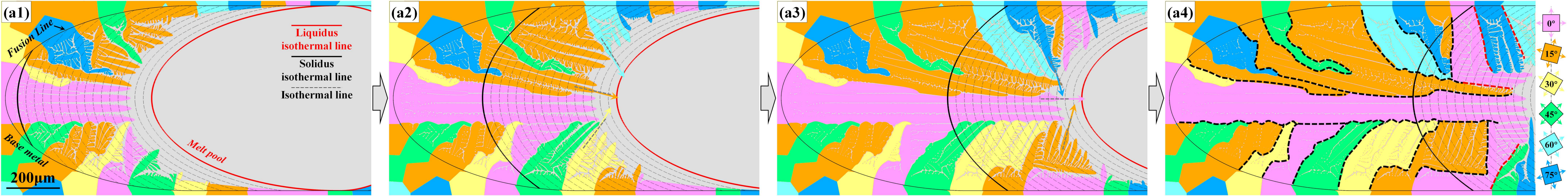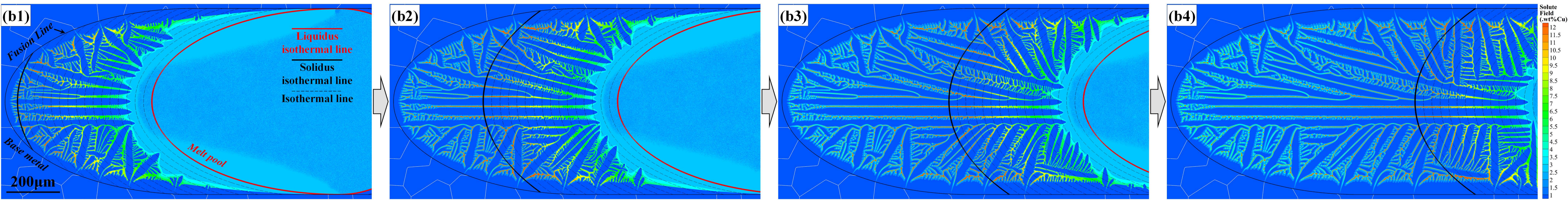